\documentclass[final,5p,times,twocolumn]{elsarticle}

\usepackage{graphicx}
\usepackage{epstopdf}

\usepackage{amssymb}
\usepackage[bottom]{footmisc}

\newcommand{\na}{$^{22}$Na}

\newcommand{\g}{$\gamma$}

\newcommand{\labr}{LaBr$_{3}$}
\newcommand{\labrs}{LaBr$_{3}$ }
\newcommand{\lacl}{LaCl$_{3}$(Ce)}  
\newcommand{\lacls}{LaCl$_{3}$(Ce) }

\newcommand{\myfigsize}{0.8}
\newcommand{\myfigsizelin}{0.4}
\newcommand{\mysinglefigsize}{0.8}

\journal{Nucl. Instr. and Meth. in Phys. Res. A (Final version at https://doi.org/10.1016/j.nima.2019.03.079)/ CC-BY-NC-ND}
\begin{document}

\date{April 8, 2019}

\begin{frontmatter}

\title{\g-Ray position reconstruction in large monolithic LaCl$_{3}$(Ce) crystals with SiPM readout}

\author{V.~Babiano, L.~Caballero, D.~Calvo, I.~Ladarescu, P.~Olleros$^1$, C.~Domingo-Pardo$^*$}
\address{Instituto de F{\'\i}sica Corpuscular, CSIC-University of Valencia, Spain}


\begin{abstract}
  We report on the spatial response characterization of large \lacls monolithic crystals optically coupled to 8$\times$8 pixel silicon photomultiplier (SiPM) sensors. A systematic study has been carried out for 511~keV $\gamma$-rays using three different crystal thicknesses of 10~mm, 20~mm and 30~mm, all of them with planar geometry and a base size of 50$\times$50~mm$^2$.
In this work we investigate and compare two different approaches for the determination of the main \g-ray hit location. On one hand, methods based on the fit of an analytical model for the scintillation light distribution provide the best results in terms of linearity and field of view, with spatial resolutions close to $\sim$1~mm \textsc{fwhm}. On the other hand, position reconstruction techniques based on neural networks provide similar linearity and field-of-view, becoming the attainable spatial resolution $\sim$3~mm \textsc{fwhm}. For the third space coordinate $z$ or depth-of-interaction we have implemented an inverse linear calibration approach based on the cross-section of the measured scintillation-light distribution at a certain height. The detectors characterized in this work are intended for the development of so-called Total Energy Detectors with Compton imaging capability (i-TED), aimed at enhanced sensitivity and selectivity measurements of neutron capture cross sections via the time-of-flight (TOF) technique. 
\end{abstract}


\end{frontmatter}

\section{Introduction}\label{sec:introduction}
In the framework of the HYMNS (High-sensitivitY Measurements of key stellar Nucleo-Synthesis reactions) project~\cite{hymns} we are developing radiation detectors with gamma-ray imaging capability aimed for demonstrating a novel technique~\cite{Domingo16,Perez16} for time-of-flight (TOF) neutron-capture cross-section measurements. The proposed detection system is based on the combination of several position-sensitive radiation detectors (PSDs) with sufficiently fast time response and good energy resolution for enabling both neutron-TOF and \g-ray Compton imaging techniques simultaneously. Thus, a set-up of two or more PSDs is operated in time-coincidence mode and arranged into a high-efficiency Compton imaging set-up, so called i-TED (Total Energy Detector with \g-ray imaging capability). Both the \g-ray imaging capability and the energy resolution are expected to provide a significant improvement in sensitivity and selectivity for true capture events with respect to commonly used systems~\cite{Guerrero13}, as described in Ref.~\cite{Domingo16}. For the implementation of the Compton technique in i-TED one needs high precision on both energy and position of the measured $\gamma$-ray interactions. In order to achieve this, the present i-TED design~\cite{Domingo16} comprises PSDs based on large monolithic \lacl-crystals optically coupled to pixelated silicon photomultipliers (SiPMs). In a previous recent publication~\cite{Olleros18} we investigated in detail the spectroscopic performance of the PSDs.\\
\rule{3cm}{0.1pt}\\
{\footnotesize
\hspace{-0.3cm}$^*$ Corresponding author.\newline
\hspace{0.2cm}E-mail address: domingo@ific.uv.es (C. Domingo-Pardo).\newline
$^1$ Present address: IMDEA Nanociencia, Campus de Cantoblanco, Spain.\newline
}
\newline
In this work we evaluate the performance of several algorithms to reconstruct the 3D-coordinates for the main \g-ray hit in the scintillation crystal.

There exist many position reconstruction algorithms for monolithic crystals available in the literature. First approaches for 2D position reconstruction, such as the centroid or Anger-logic technique~\cite{Anger58,Anger66}, use the mean value of the charge distribution collected on the photosensor (or an array of sensors) in order to infer the position of the main \g-ray hit in the transversal XY-plane of the scintillation crystal. This approach is commonly implemented by means of a resistor-network. The latter can also provide a certain sensitivity for the third spatial coordinate ($z$) or depth-of-interaction (DoI)~\cite{Lerche05}. In continuous scintillation crystals the centroid-approach works well only in the central region of the PSD, where the collected charge distribution is still rather symmetric and reflection effects in the crystal-walls have a small influence. However, strong compression or ``pin-cushion'' effects take place in the peripheral region of the PSD, thus severely reducing the FoV and linearity of the system. This can be cured, to some extent, by so-called weighted centroiding methods~\cite{Vaska01}. Also, enhanced linearity and spatial resolution have been demonstrated by the squared-charge centroiding technique reported in Ref.~\cite{Pani09}.

Over the last decade there have been many advances in terms of instrumentation for PSDs. On the one hand, the latest generation of fast and high-photon yield halide crystals~\cite{vanLoef02,Shah03,Shah05,Moses05} coupled to pixelated p-on-n semiconductor photosensors have opened up a new scope of possibilities and applications~\cite{Vaquero11,Grodzicka17}. On the other hand, the revolutionary monolithic concept~\cite{Saxena07} has eventually led to compact multi-channel photosensor-readout application-specific integrated circuits (ASICs), which enabled the possibility to build high granularity, scalable and large arrays of PSDs. For a few examples see e.g. Refs.~\cite{maroc10,peta13,DiFrancesco16}. 

Thanks to these developments, most modern position-reconstruction algorithms implement an individual multi-channel scheme for the readout of the PSDs, thereby following the concept introduced by Bird et al.~\cite{Bird94} in 1994. Presently, the former phenomenological Gaussian-based peak-fitting algorithms for position reconstruction~\cite{Truman94} have been superseded by more realistic theoretical models for the scintillation-light distribution~\cite{Lerche05,Ling08,Li10}. The latter provide indeed a better representation of the measured detector response, as reported e.g. in Refs.~\cite{Li10,Domingo09,Cabello13}. The main advantage of the analytical approach resides on the fact that, a priory, only a few parameters need to be empirically characterized. Thus, using 20$\times$20$\times$10~mm$^3$ LSO crystals coupled to 8$\times$8-pixels SiPMs spatial resolutions of 1.4~mm~\textsc{fwhm} at 511~keV have been reported~\cite{Li10}.
However, due to the theoretical nature of this methodology, experimental set-up particularities such as imperfections in the finishing of the PSDs, inhomogeneities in the crystal or in the optical coupling to the photosensor, fluctuations in the gain response for different channels, etc are not directly taken into account.
In addition, analytical methods might represent a limitation for applications requiring a real-time position reconstruction due to the relatively lengthy minimization process. Nevertheless, this constraint does not apply to neutron capture experiments where normally an offline analysis of the capture data is carried out.

Experimental details in the PSD response may be more reliably accounted for by means of an exhaustive characterization of the spatial detector response for all possible \g-ray interaction positions. This kind of pattern-shape analysis has been implemented with a great level of detail by means of statistical algorithms such as Maximum-Likelihood methods~\cite{Lerche09,Pierce18} and the so-called $k$-NN technique~\cite{Maas06,VanDam11,Aldawood17}. The latter are based on a large database of measured 2D-reference patterns, which are then used in the position-reconstruction algorithm to determine the 3D-location of the main $\gamma$-ray hit. For 50$\times$50$\times$30~mm$^3$ \labr-crystals spatial resolutions of 4.5~mm \textsc{fwhm} at 662~keV have been reported~\cite{Liprandi17} for the so-called ``Categorical Average Pattern'' extension of the aforementioned $k$-NN algorithm.

Finally, progress on computing power has also enabled the possibility to apply machine-learning artificial neural-network (NN) algorithms to the problem of the position reconstruction~\cite{Bruyndonckx07,Mateo09,Gostojic16,Conde16,Ulyanov17}. For the NN-methodology also a large database of detector responses is required, either simulated~\cite{Gostojic16} or experimentally determined~\cite{Ulyanov17}, in order to train and test the network. For example, using NNs resolutions of $\sim$2.9~mm \textsc{fwhm} and $\sim$8~mm~\textsc{fwhm} are reported for 25$\times$25$\times$10~mm$^3$ CeBr$_3$ and 28$\times$28$\times$20~mm$^3$ \labrs crystals, respectively~\cite{Ulyanov17}. In both cases SiPMs of 4$\times$4 pixels were used. Using LaBr$_3$ crystals of 50$\times$50$\times$10~mm$^3$ volume coupled to 8$\times$8 multi-anode photomultiplier-tubes (PMTs) resolutions of $\gtrsim$2~mm~\textsc{rms} ($\gtrsim$4.7~mm \textsc{fwhm}) are reported in Ref.~\cite{Gostojic16}.

In this work we explore the applicability and performance of some of these methods to three large monolithic \lacl-crystals, with a base surface of 50$\times$50~mm$^2$ and thicknesses of 10~mm, 20~mm and 30~mm. To our knowledge, these are the largest lanthanum-halide monolithic PSDs with SiPM readout aimed at $\gamma$-ray imaging reported in the literature thus far. The pixelated SiPMs, readout- and processing-electronics together with the characterization apparatus and methodology are described in Sec.~\ref{sec:setup}. The implemented position reconstruction algorithms are reported in Sec.~\ref{sec:algorithms}. The latter section is divided in three parts. The first part (3.1) describes the performance of two common methods, namely the Anger logic~\cite{Anger58}, and the squared-charge centroiding approach~\cite{Pani09}.  These basic methods have been implemented in this work with the purpose of defining the main performance figures of merit, such as resolution, linearity, field of view (FoV) and signal-to-noise (S/N) ratio. Due to their simplicity, they are still among the fastest algorithms for online monitoring during data taking and thus, we use them in this work to benchmark the speed-capability of more sophisticated approaches reported in the subsequent sections. Thus, the second sub-section (3.2) describes the performance of state-of-the-art analytical models for the propagation of the scintillation photon field within the crystal, applied to the position reconstruction along the transversal $xy$-plane. Hereby the simple model by Lerche et al.~\cite{Lerche05} is compared against the more elaborated model by Li et al.~\cite{Li10}. In Section 3.3 we describe the implementation and performance of NN-algorithms, also constrained to the transversal crystal plane. As reported in Sec.~\ref{sec:doi} we have found better results by decoupling the transversal position reconstruction (either with analytical or NN-methods), from the reconstruction in $z$ or DoI. The latter section thus describes the methodology implemented here for the reconstruction of the third space coordinate. A general comparison summarizing the advantages and drawbacks of each method is reported in Sec.~\ref{sec:summary}.

\section{Apparatus and experimental set-up}\label{sec:setup}

\subsection{\g-Ray position sensitive detectors (PSDs)}
All \lacl crystals are encapsulated in a 0.5~mm thick aluminum housing, which is isolated from the crystal itself with a 1~mm thick seam gum. The base of the crystal is optically coupled to a fused-silica glass window of 2~mm thickness. Apart from the polished base surface of the crystal, the other five surfaces are ground finished and covered with diffuse polytetrafluoroethylene reflector (PTFE) to optimize photon-collection.

Each scintillation crystal is optically coupled with silicon grease (BC-630) to a silicon photomultiplier array (SiPM) from SensL (ArrayJ-60035-65P-PCB). This sensor has a size of 50.44$\times$50.44~mm$^2$ and features 8$\times$8 pixels, each one with a size of 6$\times$6~mm$^2$ on a pitch of 6.33~mm. Each pixel features 22292 avalanche photodiodes (APDs) or micro-cells (35~$\mu$m size) and the fill-factor is of 75\%. These APDs are built using a p-on-n semiconductor structure, thus featuring the maximum of the photodetection efficiency at relatively low photon wavelengths ($\sim$420~nm), which still matches reasonably well with the main emission wavelength of 350~nm for \lacl. For more information about the SiPM the reader is referred to Ref.\cite{sensl}. For further details about the energy resolution and spectroscopic performance of these PSDs the reader is referred to Ref.~\cite{Olleros18}.

\begin{figure}[htbp!]
\flushleft
\centering
\includegraphics[width=0.95\columnwidth]{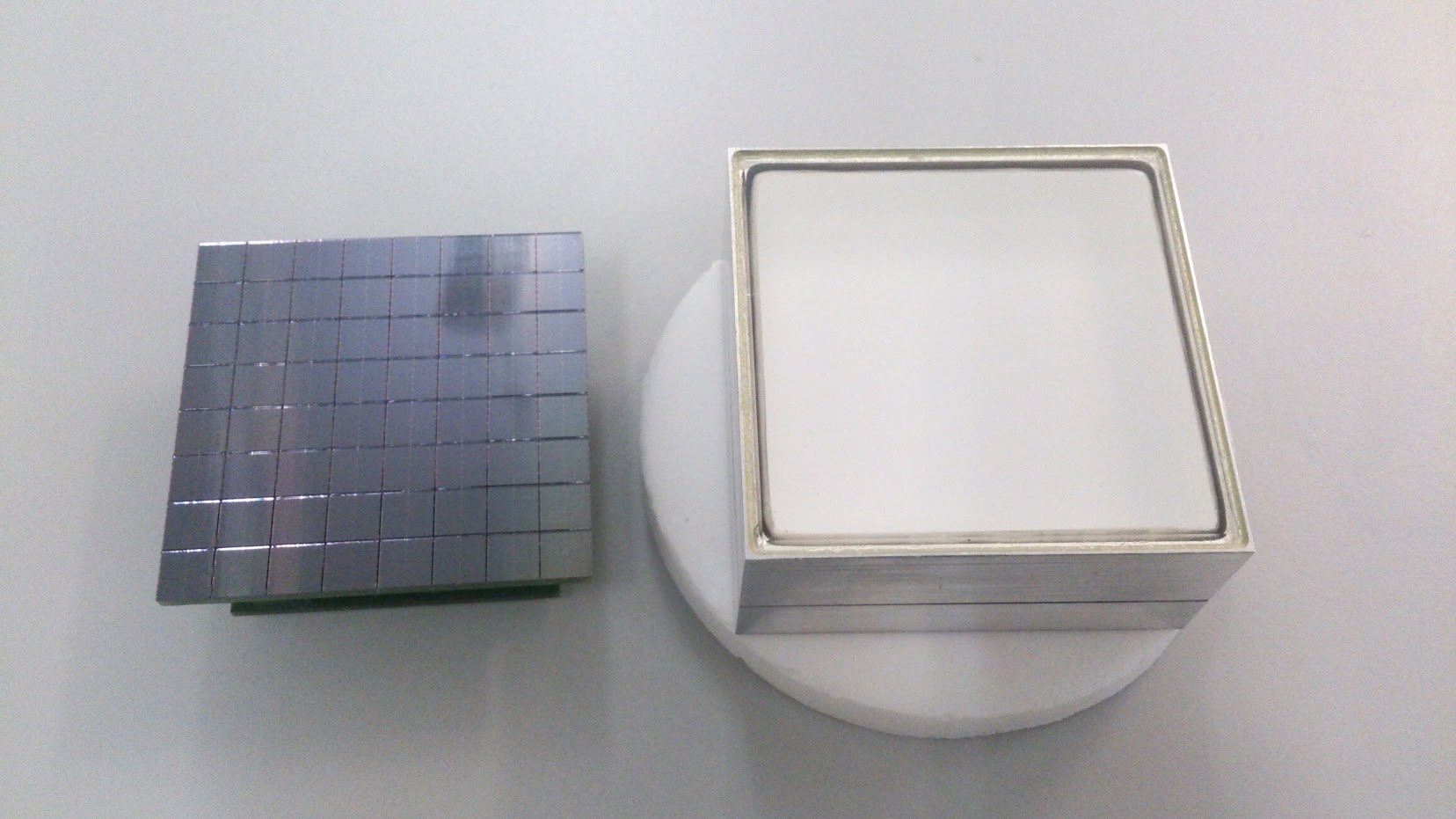}
\caption{\label{fig:crystals} Photograph of the 20~mm thick \lacls crystal together with the 8$\times$8 pixel SensL SiPM.}
\end{figure}

\subsection{Frontend SiPM readout and bias electronics}
The SiPMs are biased and readout by means of the PETsys TOF Front-End Board D version 2 (FEB/D-1024)~\cite{petsys}. The chosen reverse bias voltage is of +5~V beyond the nominal breakdown value of 24.5~V, corresponding to a nominal quantum efficiency of nearly 50\%/microcell. For the present measurements we use two of the eight acquisition ports available at the motherboard FEB/D\_v2, each port capable of acquiring up to 128 individual SiPM channels. The analogue signals are readout via 64-channel frontend ASICs (TOFPET2), which are plugged by means of a customized PCB-board to the Samtec 80-way connectors (QTE-040-03-F-D-A) at the rear-side PCB of each SiPM.
The ASIC performs the readout and digitization of the SiPM signals and uses a low threshold for timing and a high threshold for accepting the event. The maximum input dynamic range is of 1500~pC per channel.  Every time one of the 64 channels exceeds the high threshold a record is created giving the channel number, the time and the charge of the event. Digitized events contain the signal integrated charge and time stamp, and are sent via 50~cm long Samtec EQCD High-Speed flat cables to the FEB/D\_v2 motherboard, where a Kintex7 FPGA performs further event pre-processing. The communications-mezzanine sends the processed data (time-stamp, qdc and pixel identification number) via a fast Gigabit ethernet link to the acquisition computer running the system-control, bias and online monitoring software. Data are stored in binary files with a convenient format, for posterior event-building and time-coincident event selection.

The TOFPET2 ASIC features an on-chip calibration circuitry, which is used to calibrate the discriminators, TDC and QDC for all the 64 input channels. In particular, the QDC calibration is accomplished by varying the duration of the integration window for a systematic scan with test pulses provided by the FPGA. Thus, an offset current is determined for each channel, which is then removed when integrating the charge of each event.

Both the SiPM and the TOFPET2 chips are sensitive to temperature variations and thus, a system was implemented in order to keep stable thermal conditions and to constantly monitor the temperature. To this aim cooled and compressed air (1.3~bar) is constantly flushed onto the ASIC surface by means of a customized encapsulation and a tube with a diameter of 4.5~mm pointing to the ASIC. The air is cooled using a system build with four Peltier cells connected to a thermally isolated aluminum case. The hot Peltier surface is thermally coupled to a 1~cm thick heatsink block of aluminum, which is refrigerated by means of a water assisted cooling system (Kraken x52 by \textsc{nzxt}). Temperature is monitored with a $\pm$0.1~K accuracy at different points using 100K thermistors connected to the main acquisition computer through an Arduino controller board. Temperature is monitored inside the cooler-case, at the hot area of the aluminum block and inside each detector housing. In such locations, the typical stable temperature during acquisition is 12$^{\circ}$C, 50$^{\circ}$C and 21.5$^{\circ}$C, respectively.


\subsection{Scanning table and data-sets}
For the systematic scan of the full detector surface of 50$\times$50~mm$^2$ we use a refurbished version of an XY table from \textsc{arrick-robotics}~\cite{arrick} equipped with a low-stretch timing belt and stepper motors. Each of the two stepper motors for X and Y positioning is connected to a 2.5:1 pulley reducer, which enables a positioning resolution of about 80~$\mu$m/step and a repeatability of 0.2~mm. This accuracy was checked by means of a digital microscope and a calibration slide. The XY-positioning is synchronized with our acquisition system (see below) in order to trigger and stop data-taking and to store data-files with proper names indicative for the scanning position of each acquisition. To this aim a software code was written, which reads a user-provided configuration file with a series of position coordinates $x$ and $y$, and the preset acquisition time for each scan position. 

\begin{figure}[htbp!]
\flushleft
\centering
\includegraphics[width=0.5\columnwidth]{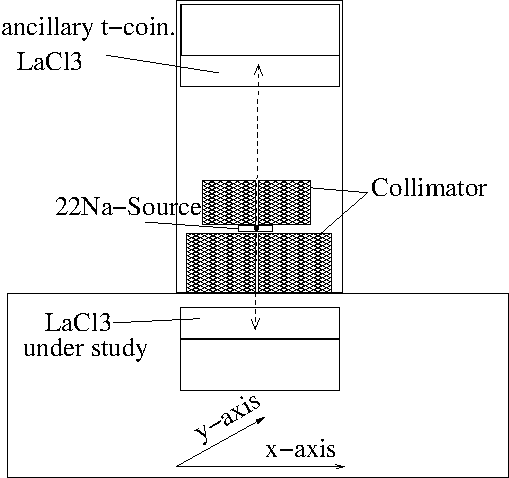}
\vfill
\includegraphics[width=\columnwidth]{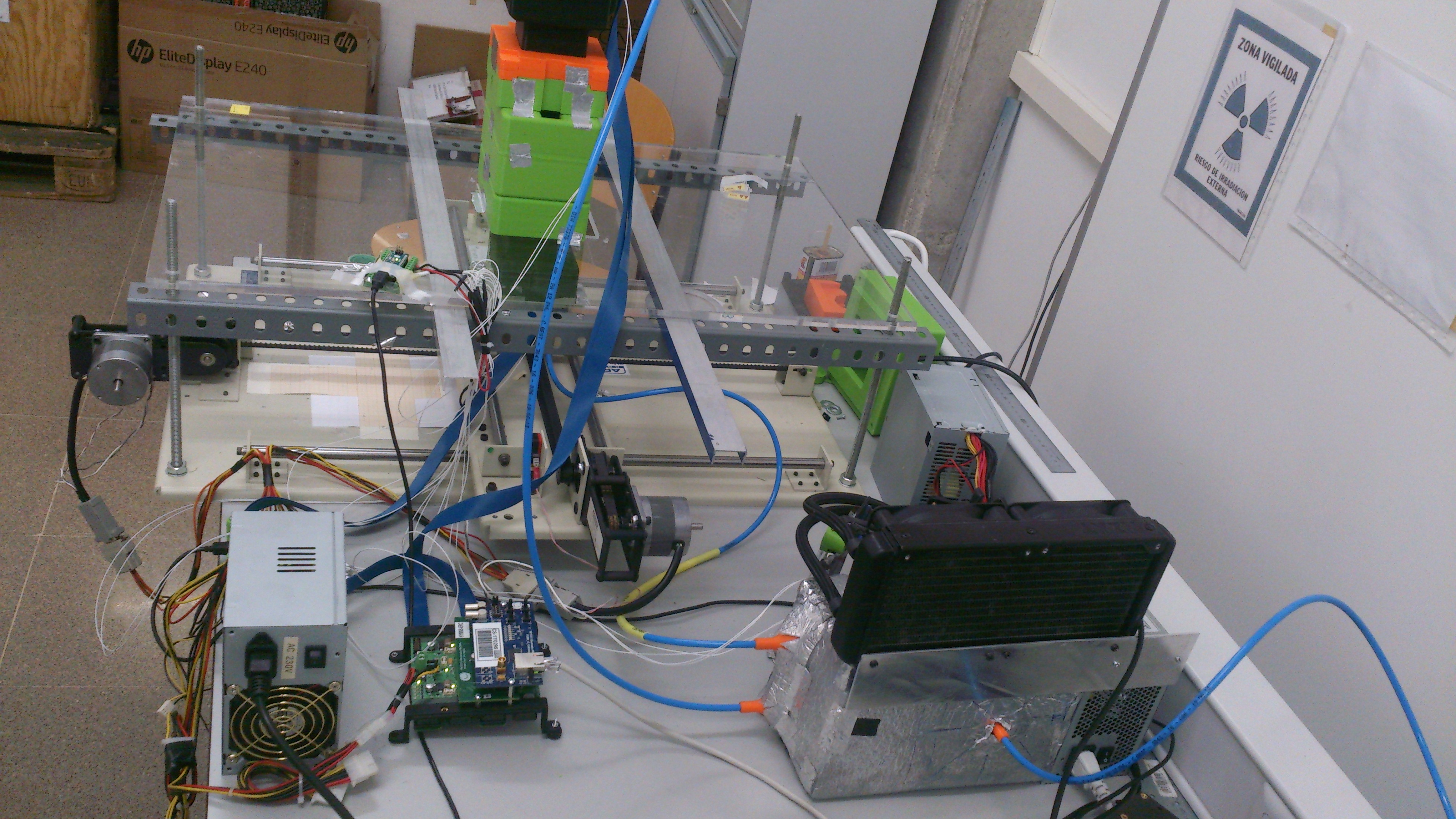}
\caption{\label{fig:setup} Characterization workbench with the XY scanning-table. The two \lacls detectors are mounted in vertical position, as illustrated in the schematic drawing. The detector at the top remains fixed and coupled to the collimated $^{22}$Na positron source. The bottom detector sits on the movable XY-bench for characterization.}
\end{figure}

In order to suppress the self-activity of the \lacls scintillation crystals all measurements were made in time-coincidence between two detectors using the 511~keV annihilation gamma-quanta emitted isotropically in back-to-back direction from a point-like \na-source with an activity of 416~kBq. The coincidence time-window was set to 20~ns. The PSD under characterization was attached to a small movable platform, whereas the collimator, the \na-source and the ancillary detector were fixed on top of a 10~mm thick platform from Plexiglas (see Fig.~\ref{fig:setup}). The collimator is a parallelepiped made from tungsten, with a central hole-diameter of 1~mm and a thickness of 30~mm.

\begin{figure}[htbp!]
\flushleft
\centering
\includegraphics[width=0.65\columnwidth]{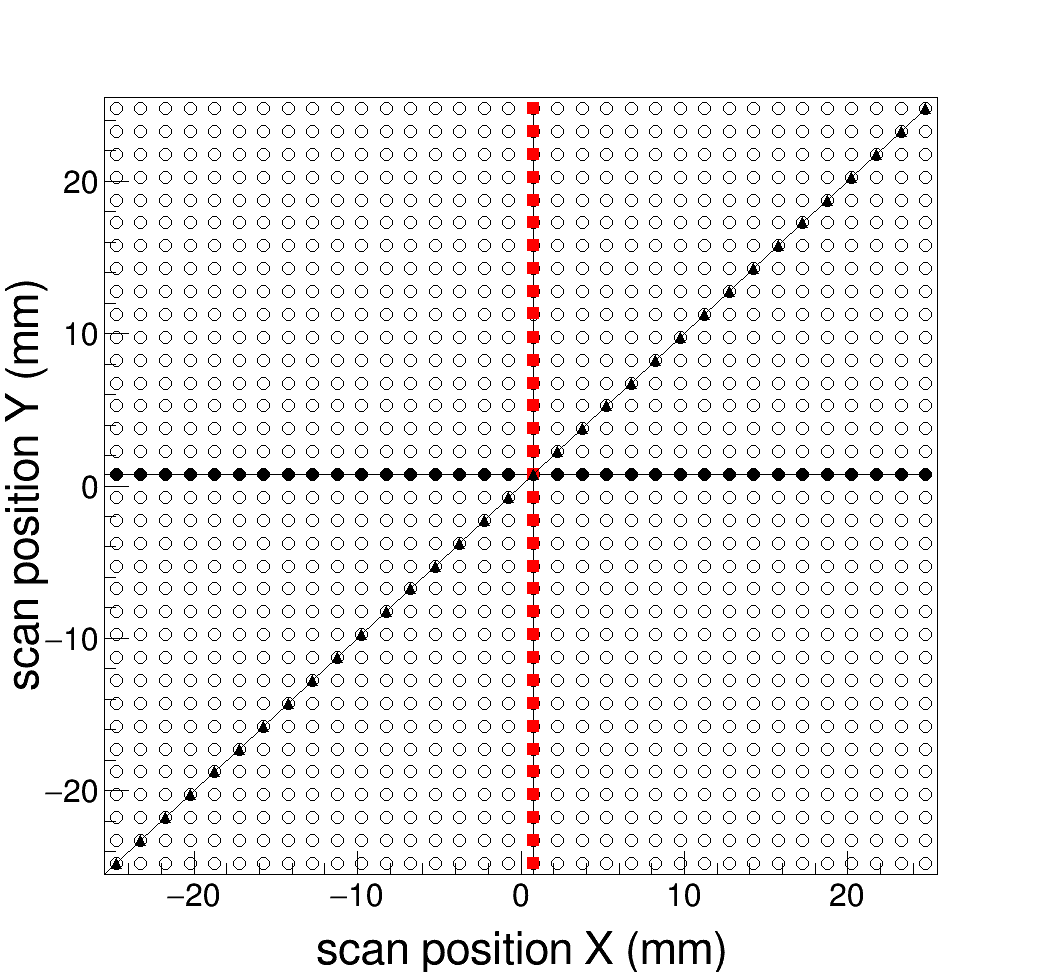}
\caption{\label{fig:scan_sketch} Schematic drawing of the 35$\times$35 scan positions on a pitch of 1.5~mm (open circles). Solid symbols are displayed for the proper interpretation of the linearity curves described in the sections below.}
\end{figure}

A total of three data-sets were acquired, one for each crystal. Each data-set is composed of a matrix of 35$\times$35 collimated positions on a grid with pitch of 1.5(1)~mm. The scanned positions are schematically represented in Fig.~\ref{fig:scan_sketch}. Data were acquired for each scan position during a time interval of 600~s. Thus, the scan of each detector lasted for about 8 days.

Hereafter, the 35 collimated positions along central $x-$axis of the PSD (solid black circles in Fig.\ref{fig:scan_sketch}) and the 35 positions along the orthogonal $y-$axis (solid red circles) are referred to as horizontal and vertical scanning lines or central cross. The 35 collimated positions (solid triangles) depicting a 45$^{\circ}$ line with respect to the previous two directions are referred to as diagonal scanning line.

\subsection{Deconvolution of the collimated \g-ray beam divergence}
\begin{figure}[htbp!]
\flushleft
\centering
\includegraphics[width=\mysinglefigsize\columnwidth]{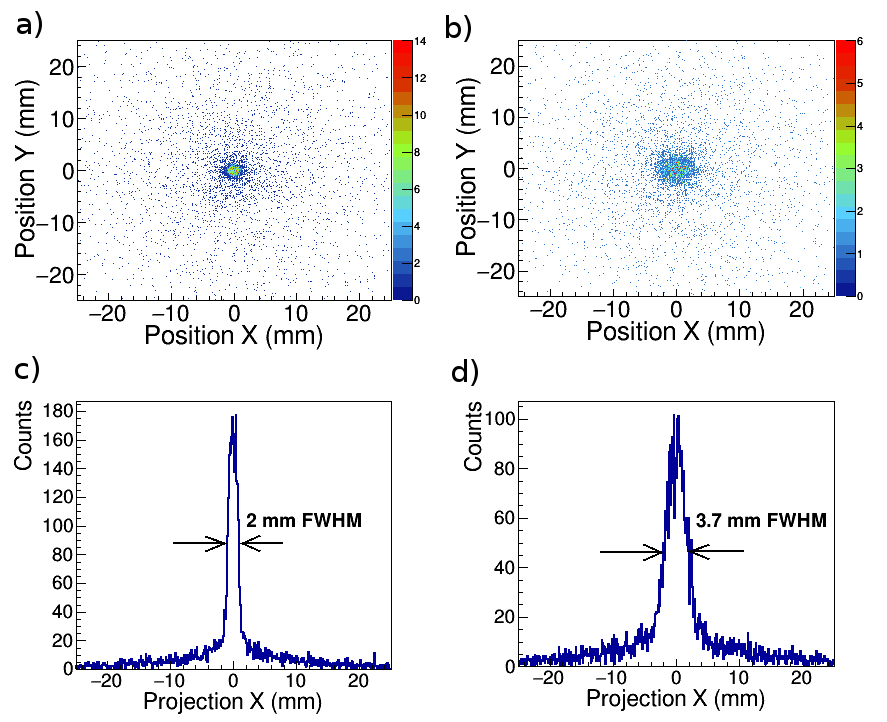}

\caption{\label{fig:deconv20mm} MC simulation for 511~keV \g-rays passing through the 1~mm diameter collimator and impinging at the center of the 20~mm thick \lacls detector (see text for details). Figures a) and b) show the simulated 2D-distribution for an ideal detector and for a detector with an intrinsic resolution of 3~mm \textsc{fwhm}, respectively.  The bottom panels c) and d) show the corresponding projections over the $x$-axis, thus yielding an intrinsic set-up related broadening of 2~mm \textsc{fhwm} and an overall broadening (detector/algorithm plus set-up) of 3.7~mm \textsc{fwhm}.}

\end{figure}

In order to determine the intrinsic detector spatial resolution from a measurement made with a collimated $\gamma$-ray source it becomes necessary to deconvolute the spatial spread. Apart from the intrinsic resolution related to the detector and to the reconstruction algorithm itself, the overall broadening is also affected by the beam divergence originating from the collimator aperture, the thickness and the distance to the detector under study. The latter contribution has been quantified by means of Monte Carlo (MC) simulations using the \textsc{Geant4} code~\cite{geant4}. The experimental set-up was included in the simulation and special care was taken to model in a realistic way all sensible distances and materials. For each crystal thickness a total of 1$\times$10$^9$ events from an isotropic source of 511~keV \g-rays were simulated. One example for the position spread in the 20~mm thick crystal is shown in Fig.~\ref{fig:deconv20mm}.

\begin{figure}[htbp!]
\flushleft
\centering
\includegraphics[width=\mysinglefigsize\columnwidth]{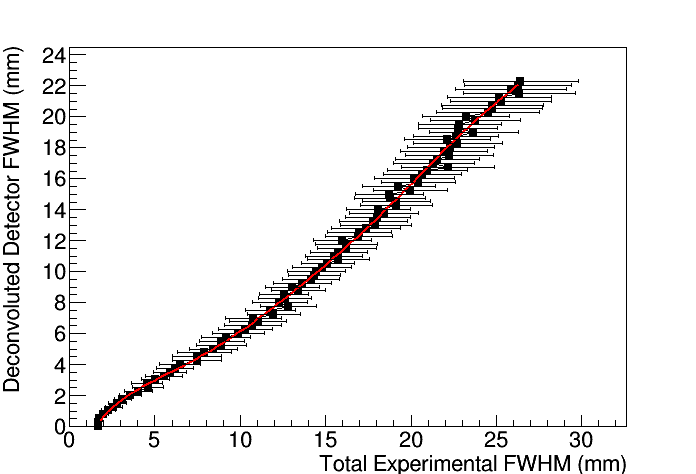}
\caption{\label{fig:decfuncs} Deconvolution function obtained from the MC-simulated and convoluted responses for the 20~mm thick \lacls crystal.}
\end{figure}

To determine the relation between the ``true'' or intrinsic detector spatial resolution and the total or ``measurable'' width, for each simulation we convolute the simulated positions of the \g-ray hits with a Gaussian function. This convolution is carried out for a series of Gaussian widths spanning from 0~mm up to 22~mm \textsc{fwhm}. The result for the 20~mm thick crystal is displayed in Fig.~\ref{fig:decfuncs}, which shows the dependency between the convolution width or instrumental resolution versus the overall (divergence affected) distribution width. For convenience the simulated data-points are adjusted to an arbitrary polynomial function, which is then used along this work in order to derive the intrinsic detector resolution from the total measured width.

For the sake of clarity, results are exemplified here only for the 20~mm thick \lacls crystal. Similar calculations were also carried out for the other two crystal thicknesses of 10~mm and 30~mm and applied consequently along this work for the deconvolution the intrinsic resolution associated to the detector and the reconstruction algorithm.

\section{$\gamma$-Ray position reconstruction algorithms}\label{sec:algorithms}
This section is structured as follows. Sec.~\ref{sec:centroid} describes briefly two of the most common reconstruction techniques, namely the Anger-logic method~\cite{Anger58,Anger66} and a variation of it, the so-called squared-charge centroiding approach~\cite{Pani09}. The performance of these algorithms is in actuality superseded by other techniques, such as those reported in the subsequent sections. Nevertheless we have implemented via software these two basic position-reconstruction methods in order to define the main performance figures of merit used along this work. They also serve to illustrate the improvement attained with more advanced apparatus and analysis approaches. On the other hand, owing to the simplicity of these methods, they are the fastest from the computational point of view and thus, serve as reference for benchmarking the computational load of the other methods. State-of-the-art algorithms like those based on analytical models and artificial NNs are reported in Sec.~\ref{sec:fit} and Sec.~\ref{sec:nn}, respectively. For all algorithms only events with a full-energy deposition are taken into account. Unless otherwise stated, a flood-illumination measurement for each crystal/SiPM assembly was used to correct for pixel-gain fluctuations. Such corrections were applied on an event-by-event basis before the position-reconstruction analysis.

\subsection{Anger-logic and squared-charge techniques}\label{sec:centroid}

The Anger-technique is based on the use of a resistor network~\cite{Anger58,Anger66} coupled to an array of phototubes (SiPM-pixels in our case). The pulse-height of the electrical signal measured at each one of the four network corners becomes proportional to the gamma-ray hit distance. The location coordinates are then determined by using the Anger formula (see for example Eq.(2a) and Eq.(2b) in Ref.\cite{Anger66}). 
In order to emulate the centroiding method, instead of implementing it by hardware, we have followed a software approach. The latter is based on the computation of the mean-value of the charge-distributions measured with our pixelated SiPMs. For the squared-centroiding method~\cite{Pani09} the mean value of the squared-charge distribution is used instead. Thus, the coordinates of the reconstructed position ($r_{x}, r_{y}$) for any registered event can be computed as follows

\begin{equation}
  r_{k} = \frac{\sum_{i=0}^{63}q^m_i r_{k,i}}{\sum_{i=0}^{63}q^m_i}, 
\end{equation}

where $k=x,y$, $m=1$ for Anger-logic or $m=2$ for the squared-charge centroiding technique, $(r_{x,i}$ and $r_{y,i})$ represent the $x$ and $y$ coordinates for pixel $i$ containing a total charge $q_i$ (a.u.).

Using a resistor network it is not possible to select a different number of channels for each registered event. Therefore, all 64 pixels available were included in our software approach for the centroiding position reconstruction.
As illustrative reference, position-reconstruction examples for the central scan position of true coordinates ($x_{true}=0~mm, y_{true}=0~mm$) and a peripheral scan position ($x_{true}=21~mm, y_{true}=21~mm$) acquired for the 20~mm thick \lacls crystal are shown below in Fig.~\ref{fig:centroid_examples} and Fig.~\ref{fig:pani} for both algorithms. The enhancement in FoV obtained with the second approach becomes directly apparent when comparing the peripheral-position distributions (panels c) and d) in Fig.\ref{fig:centroid_examples} and Fig.\ref{fig:pani}).

\begin{figure}[htbp!]
\flushleft
\centering
\includegraphics[width=\myfigsize\columnwidth]{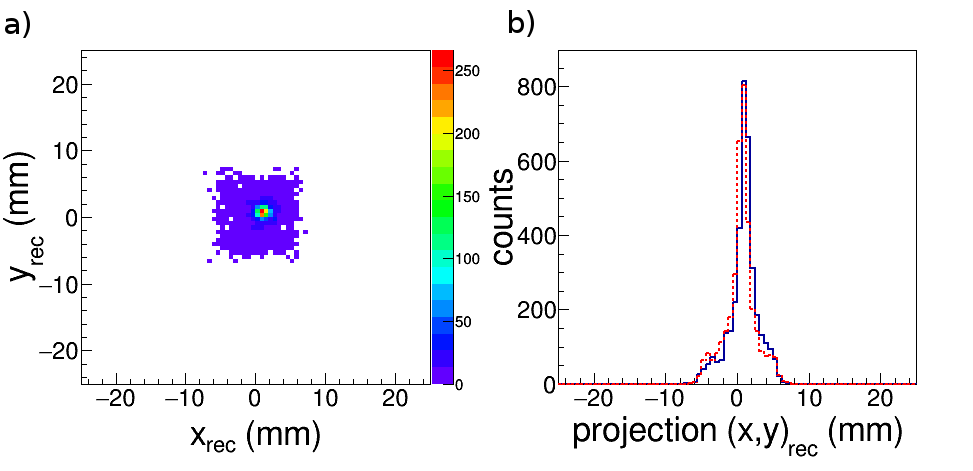}
\includegraphics[width=\myfigsize\columnwidth]{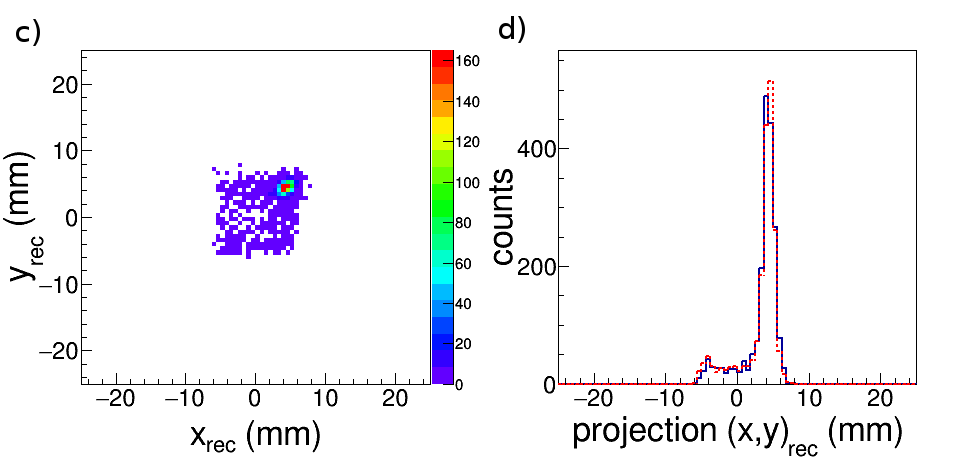}
\caption{\label{fig:centroid_examples} Event-by-event reconstructed 2D-locations using a software implementation of the Anger-logic technique for the central scan position (a). Panel (b) shows the projected position distribution over the $x$-axis (solid line)  and over the $y$-axis (dashed line). Panels (c) and (d) show equivalent distributions for a scan position shifted 21~mm in $x$ and $y$ with respect to the center, i.e towards the top-right crystal corner.}
\end{figure}

\begin{figure}[htbp!]
\flushleft
\centering
\includegraphics[width=\myfigsize\columnwidth]{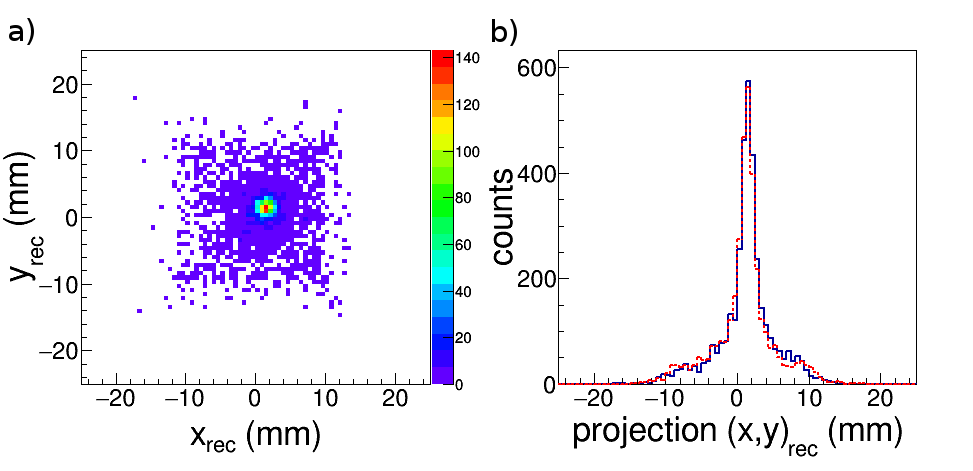}
\includegraphics[width=\myfigsize\columnwidth]{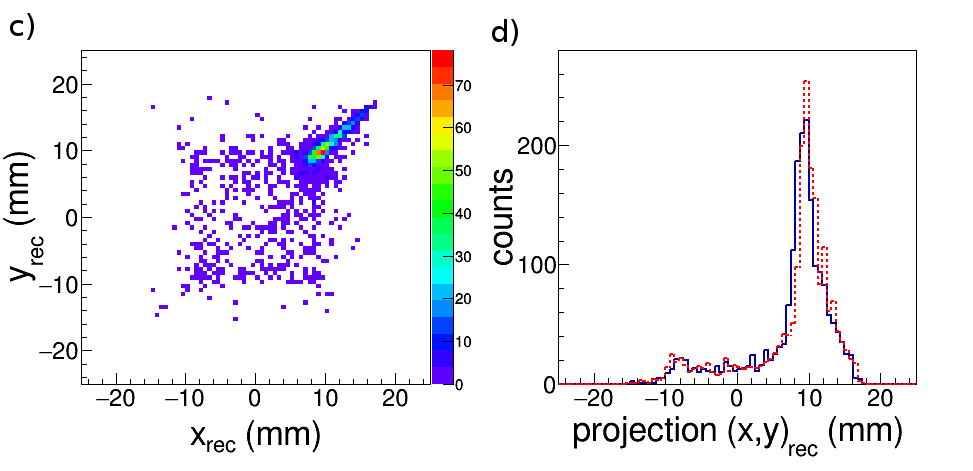}
\caption{\label{fig:pani} Same as Fig.~\ref{fig:centroid_examples} using a software approach for the squared-charge centroiding method~\cite{Pani09}.}
\end{figure}

\begin{figure}[htbp!]
\flushleft
\centering
\includegraphics[width=\myfigsizelin\columnwidth]{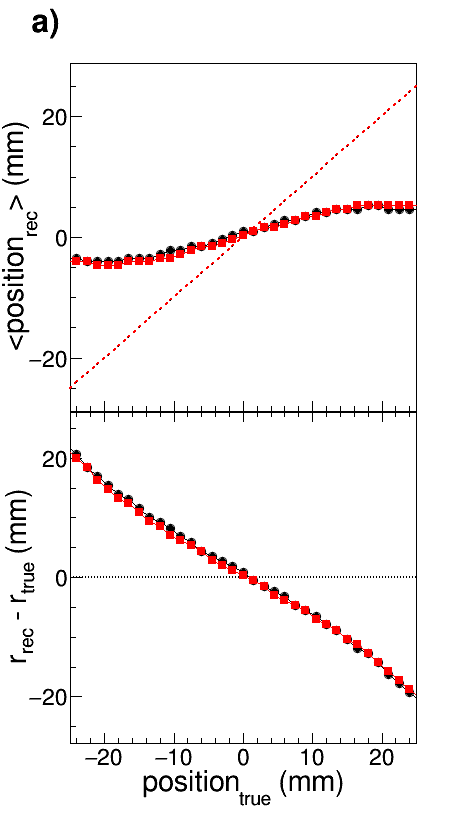}
\includegraphics[width=\myfigsizelin\columnwidth]{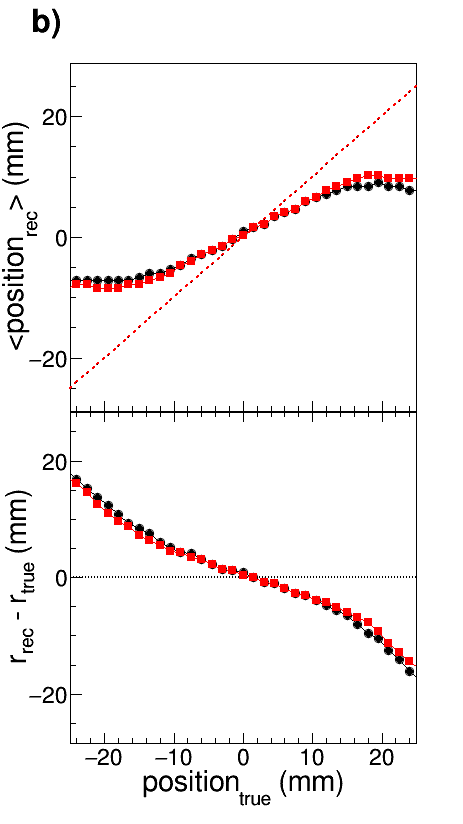}
\caption{\label{fig:centroid} Linearity  obtained with the Anger-logic technique (a) and the squared-charge method (b) for scan positions along the central $x$-axis (black circles) and $y$-axis (red squares) of the crystal.}
\end{figure}

\begin{figure}[htbp!]
\flushleft
\centering
\includegraphics[width=\mysinglefigsize\columnwidth]{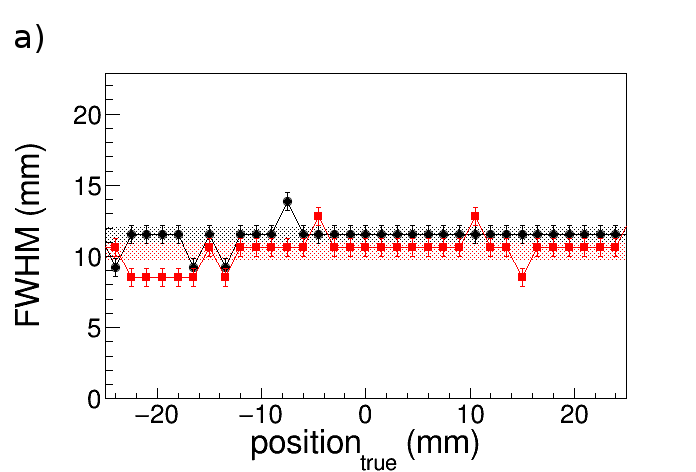}
\includegraphics[width=\mysinglefigsize\columnwidth]{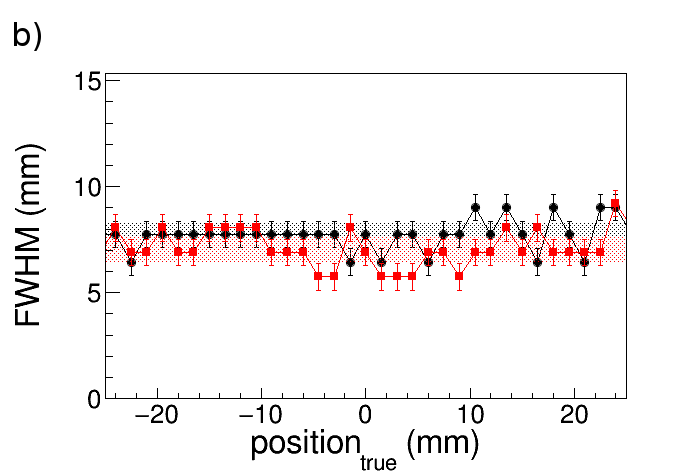}
\caption{\label{fig:centroid_fwhm} Spatial resolution \textsc{fwhm} obtained with the Anger-logic technique (a) and for the squared-charge centroiding method (b) for scan positions along the central $x$-axis (bold-circles) and $y$-axis (red-squares) of the crystal. Shaded bands represent average resolution values in the detector FoV.}
\end{figure}

The linearity curve is defined here as the relation between the mean value of the reconstructed positions and the true position for each scanned point. This quantity is shown in the top panels of Fig.~\ref{fig:centroid} for the central cross of 35-horizontal and 35-vertical scanning positions. These 69 positions (note that the central position is common to both data-sets) are those represented in Fig.\ref{fig:scan_sketch} by solid black circles and solid red squares, respectively. The thin-dashed line in the top panels shows the behavior expected for an ideal detector. On the other hand, deviations between reconstructed and true positions ($r_{rec} - r_{true}$) are displayed in the bottom panels. Thus, the slope of the linearity curve in the central linear region provides a measure of the quality of the algorithm in terms of image compression. A linearity slope of 45$^{\circ}$ corresponds to a 1:1 relationship between true and reconstructed position and hence to an ideal detector. For that ideal case deviations ($r_{rec} - r_{true}$) shown in the bottom panels would vanish.
For the Anger-logic approach the slope of the linearity curve is of only 30(1)\% in the central region. The spatial resolution, defined as the \textsc{fwhm}-value of the x (y) projected-distribution for the reconstructed positions along the x (y) axis are shown in Fig.\ref{fig:centroid_fwhm}; Along the central cross of scan positions the average resolution is of 10.8(6)~mm \textsc{fwhm}. These values include also a correction for the aforementioned linearity distortion.


A noticeable improvement in linearity and resolution is obtained simply by working with the squared-charge values of each pixel, as reported in Ref.~\cite{Pani09}. This is demonstrated in Fig.~\ref{fig:centroid}-b), which shows an improved average linearity of 51(1)\% along the $x$- and $y$-axis. The average spatial resolution is of 7.3(6)~mm~\textsc{fwhm} (Fig.\ref{fig:centroid_fwhm}-b). The spatial response is still remarkably affected by border effects. Finally, it is worth noting that the linearity curves displayed in Fig.~\ref{fig:centroid} are quite similar to the comparison reported in Fig.~3 of Ref.\cite{Pani09}, thus in agreement with the approximations of the software-approach implemented here.

The field-of-view (FoV) is defined here as the sensitive PSD surface where the linearity curve along the $x$- and $y$-axes shows a strictly increasing behaviour. From the linearity curves displayed in Fig.~\ref{fig:centroid} the FoV becomes 30$\times$30~mm$^2$ and 36$\times$36~mm$^2$ for the Anger and for the squared-charge centroiding methods, respectively. Despite of their limited performance, a clear advantage of these methods is their reconstruction speed. Using a computer with a core i7 from Intel, processing rates of $r_{Anger} = 6840$~Events/s and $r_{Q^2} = 6647$~Events/s were obtained for the Anger- and squared-charge centroiding techniques, respectively. The value of $r_{Anger}$ will be used in the rest of this work in order to benchmark the processing-speed performance of the other methods.

\subsection{Analytical model fit}\label{sec:fit}

There exist several analytical models to describe the 3D-spatial propagation of the scintillation photons produced by a single $\gamma$-ray hit or, equivalently, by a point-like photon source within the crystal volume~\cite{Lerche05,Ling08,Li10}.
Here we report on the implementation of algorithms based on both the model by Lerche et al.~\cite{Lerche05} and the somewhat more elaborated formula by Li et al.~\cite{Li10}. 
The Lerche model makes use of the inverse square law combined with an exponential factor, which accounts for photon absorption and scattering effects within the crystal. An additional constant term is used to take into account the scintillation-light or electronic-noise background. Two parameters $L_{\circ}$ and $\alpha$ account for the intensity of the photon source and for the average absorption, respectively. Using a dedicated flood-illumination measurement for each crystal we determine these two parameters empirically and fix them to their mean value.

\begin{figure}[htbp!]
\flushleft
\centering
\includegraphics[width=\myfigsize\columnwidth]{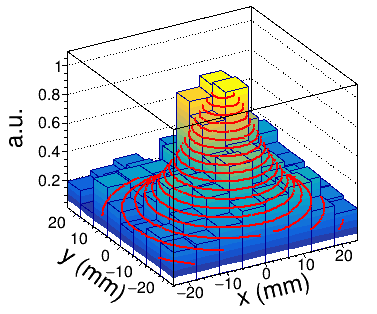}
\caption{\label{fig:chi2_10mm} Example of an arbitrary event measured with the 10~mm thick crystal and fitted to the Lerche-model.}
\end{figure}

The model by Li includes additional reflection effects at the walls of the crystal, a feature which seems convenient in our case due to the PTFE reflector used (see Sec.~\ref{sec:setup}). Additionally, a cut-off factor $\beta$ is used to describe the crossover from the refractive to the reflective regime. The exact value for this parameter has a small impact on the results and, as recommended~\cite{Li10}, we use a constant value of $\beta=100$.

To implement these algorithms, the PSD charge-response measured with the SiPM is stored on an event-by-event basis in a 2D-histogram, which is then fitted to the corresponding formula (Lerche or Li) using the log-likelihood method.  For this we make use of the \verb+TMinuit+ minimization class of the CERN ROOT package~\cite{root}. An example for the Lerche-model fit is displayed in Fig.~\ref{fig:chi2_10mm}. The analytical fit method allows one to use the resulting $\chi^2$-value, on an event-by-event basis, in order to reject events where the model is not reproducing well the measured charge distribution. This feature becomes helpful for the analysis of thick scintillation crystals, as it is demonstrated and outlined below.

\subsection*{\lacls 50$\times$50$\times$10~mm$^3$}
\begin{figure}[htbp!]
\flushleft
\centering
\includegraphics[width=\myfigsize\columnwidth]{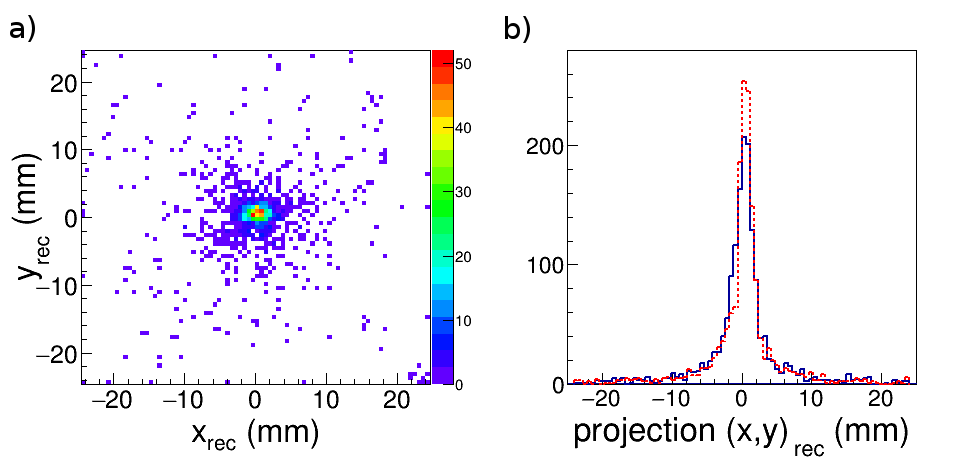}
\includegraphics[width=\myfigsize\columnwidth]{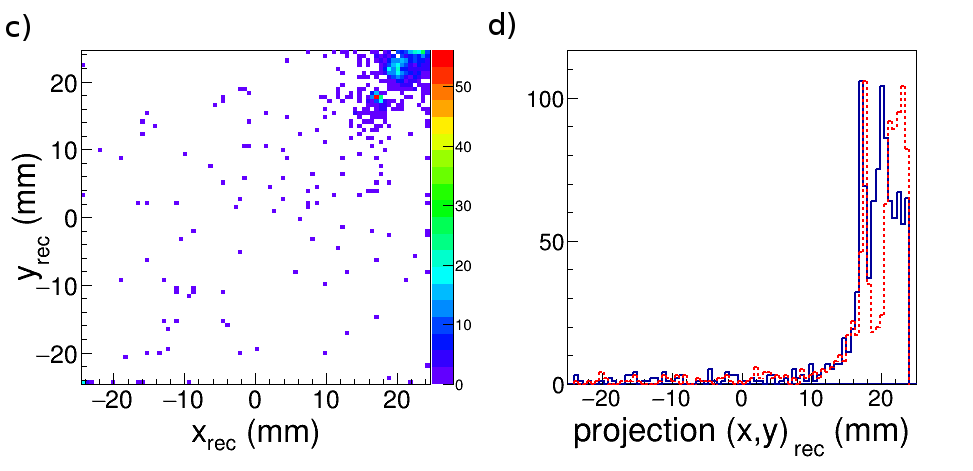}
\caption{\label{fig:examples_Lercheanalytical_10mm} Examples for central (a-b) and peripheral (c-d) scan-positions in the 10~mm thick crystal reconstructed with the Lerche-fit method.}
\end{figure}

\begin{figure}[htbp!]
\flushleft
\centering
\includegraphics[width=\myfigsize\columnwidth]{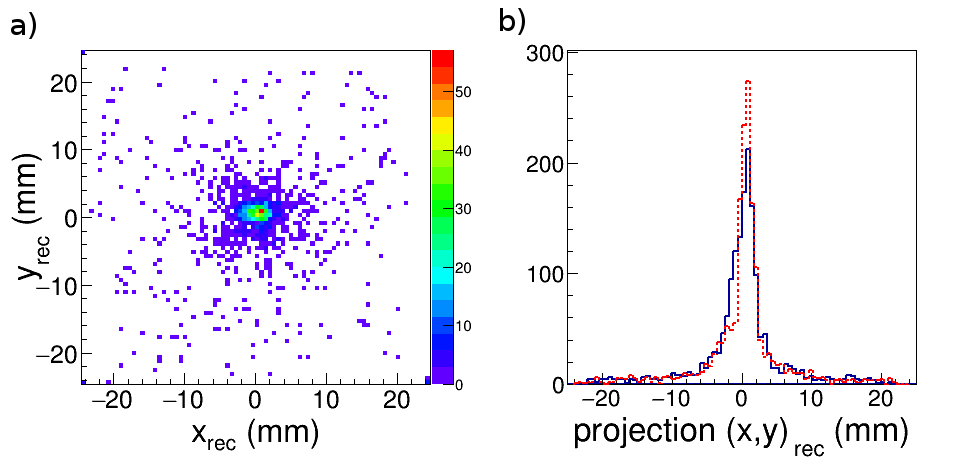}
\includegraphics[width=\myfigsize\columnwidth]{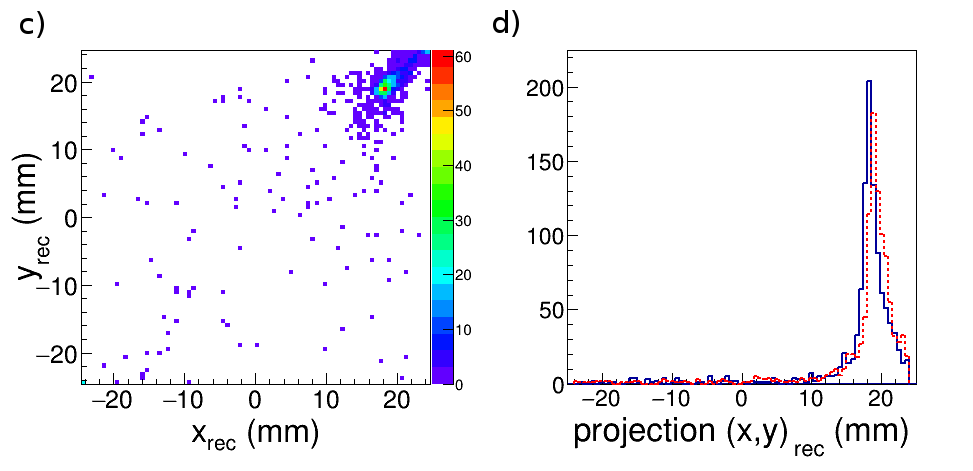}
\caption{\label{fig:examples_Lianalytical_10mm} Same examples as those shown in Fig.~\ref{fig:examples_Lercheanalytical_10mm} using the Li-fit method.}
\end{figure}

The reference examples of spatial distributions are shown in Fig.~\ref{fig:examples_Lercheanalytical_10mm} and Fig.~\ref{fig:examples_Lianalytical_10mm} for the Lerche and Li models, respectively. The two positions represented, ($x_{true}=0~mm, y_{true}=0~mm$) and ($x_{true}=21~mm,y_{true}=21~mm$), are the same as those shown in Fig.~\ref{fig:centroid_examples} and Fig.~\ref{fig:pani} for the centroiding approaches.

Near the corner of the crystal the Li-model shows a superior performance, as it can be observed by comparing panels c) and d) in Fig.~\ref{fig:examples_Lercheanalytical_10mm} and Fig.~\ref{fig:examples_Lianalytical_10mm}. It is worth emphasizing that the latter two figures correspond to the same data-set, being the scintillation-light model the only difference in the algorithm used for the position reconstruction.
In order to reliably quantify the size and the geometry of the FoV one needs to consider that light-reflection effects become more accute at the corners of the crystal than in the central wall region between corners. A  position-fit reconstruction algorithm can be very sensitive to such effects and, therefore, it becomes convenient to evaluate the linearity of the system not only along the central $x-$ and $y-$ axis, but also along the diagonal of the PSD (see Fig.~\ref{fig:scan_sketch}). Thus, the linearity curves for both the diagonal scan and for the central cross along the $x$- and $y$-axis are displayed below in Fig.~\ref{fig:linearity_fit_diagonal} and Fig.~\ref{fig:linearity_fit_hv}, respectively.

Using the Lerche (Li) model we find a linearity range of 39~mm (43.5~mm) and 43.5~mm (46.5~mm) along the crystal diagonal and central-cross of scanned positions, respectively (see Fig.~\ref{fig:linearity_fit_diagonal} and Fig.~\ref{fig:linearity_fit_hv}). The slightly better performance by the Li-model in the peripheral region can most probably be ascribed to the modelling of light-reflection effects. In general, we find that the FoV is constrained by the linearity performance along the diagonal direction, rather than along the central $x$ and $y$-axis. Therefore we use the diagonal scan to define the size of a squared-linear FoV. Thus, for the 10~mm thick crystal we obtain a FoV of 15.2~cm$^2$ and 18.9~cm$^2$ for the Lerche- and Li-models, respectively. Within the quoted FoV the linearity is practically 100\%, with root-mean-square (\textsc{rms}) deviations of $\lesssim$0.9~mm for both models. A summary of the main performance results is listed in Table~\ref{tab:analytic} at the end of this section.

\begin{figure}[htbp!]
\flushleft
\centering
\includegraphics[width=\myfigsizelin\columnwidth]{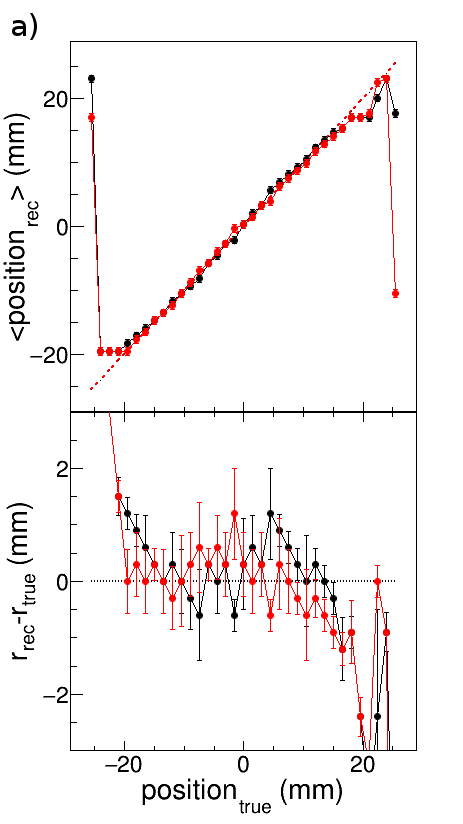}
\includegraphics[width=\myfigsizelin\columnwidth]{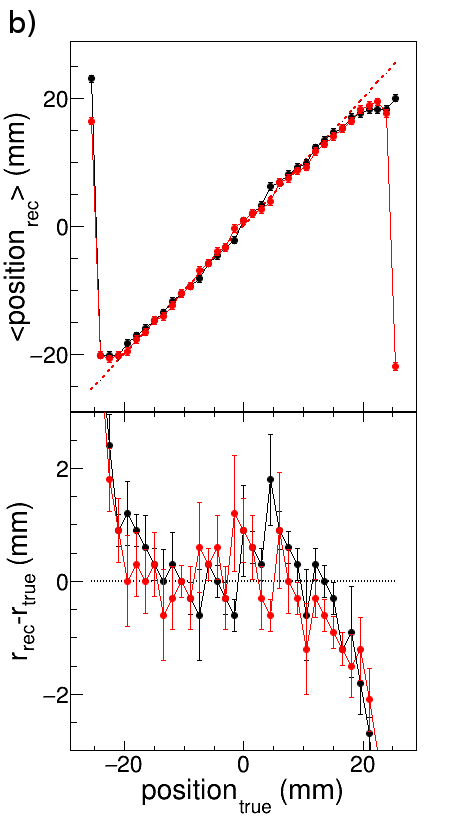}
\caption{\label{fig:linearity_fit_diagonal} Linearity along the diagonal of the 10~mm thick crystal for the Lerche-fit (a) and Li-fit (b) methods.}
\end{figure}
 
\begin{figure}[htbp!]
\flushleft
\centering
\includegraphics[width=\myfigsizelin\columnwidth]{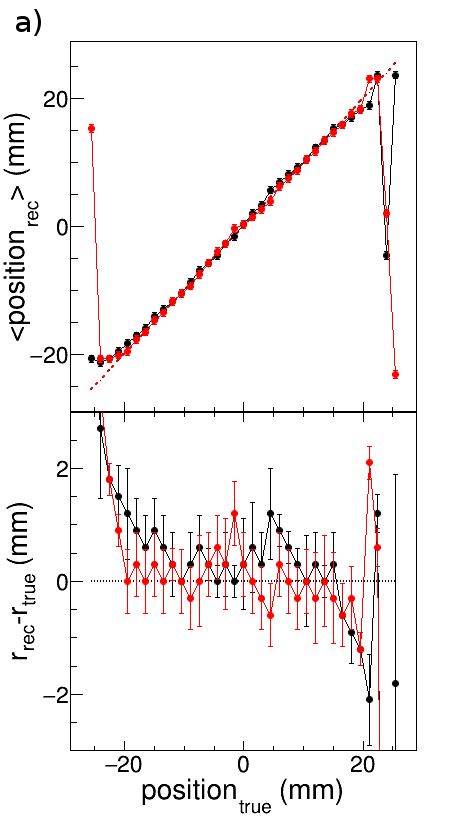}
\includegraphics[width=\myfigsizelin\columnwidth]{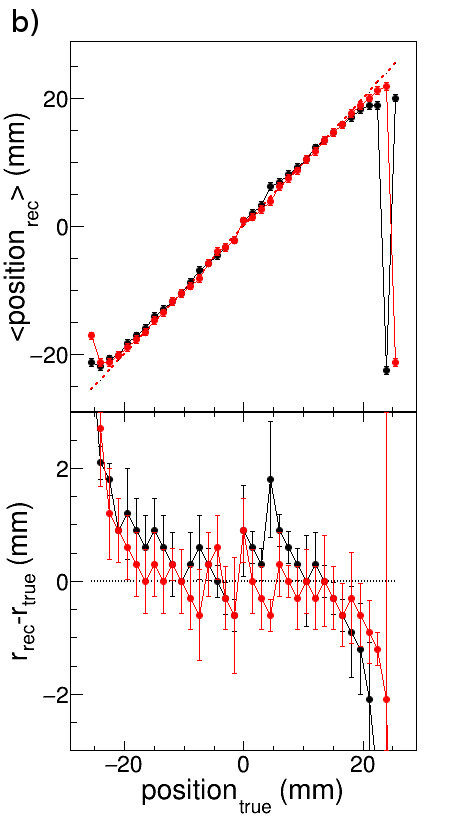}
\caption{\label{fig:linearity_fit_hv} Linearity along the horizontal ($x$) and vertical ($y$) crystal axis in the 10~mm thick crystal using the Lerche-fit (a) and Li-fit (b) methods.}
\end{figure}

In terms of spatial resolution the performance of both Lerche- and Li-models is quite similar, with average values of $\sim$1.2~mm \textsc{fwhm}, as summarized below in Table~\ref{tab:analytic}.
\begin{figure}[htbp!]
\flushleft
\centering
\includegraphics[width=\myfigsize\columnwidth]{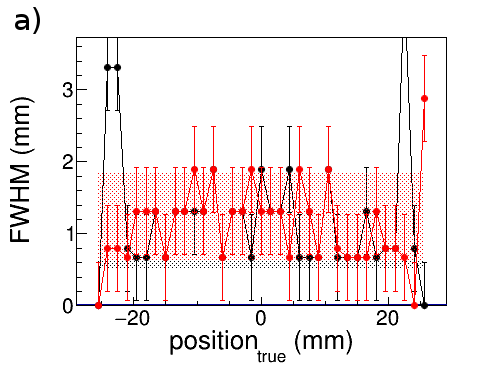}
\includegraphics[width=\myfigsize\columnwidth]{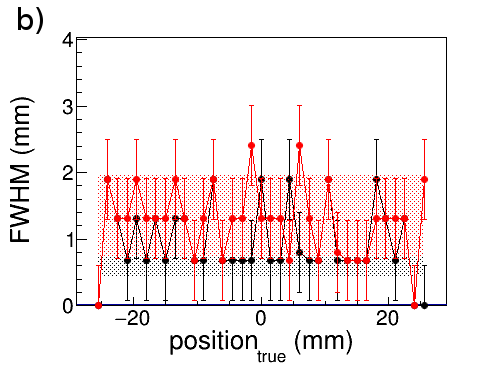}
\caption{\label{fig:resolution10mmAnlytical} Spatial resolutions \textsc{fwhm} obtained with the Lerche-model fit (a) and Li-model fit (b) for the 10~mm thick crystal.}
\end{figure}

Another important aspect to quantify the overall performance of the position-reconstruction algorithm is the S/N-ratio. For the present study we define the S/N as the density of events within full-width-at-tenth of the maximum (\textsc{fwtm}) for the 2D-distribution, normalized by density of ``stray'' events outside that region. For the 10~mm thick crystal we obtain average S/N-ratios of 14(3) and 12(5) for the Lerche and Li models, respectively (see Table~\ref{tab:analytic}).

Finally, it is worth to emphasize that the performance found here for the 10~mm thick crystal using analytical methods is rather satisfactory. Our results are comparable to those reported by other groups that have implemented also analytical methods with crystals of smaller size but similar thickness~\cite{Ling08,Li10}. Indeed, previous studies focus on relatively thin crystals ($\lesssim$10~mm), with thicknesses spanning from 10~mm~\cite{Li10,Cabello13} down to 8~mm~\cite{Ling08} and 5~mm~\cite{Domingo09,Cabello13}. Apparently, the applicability of analytical methods to crystals with thickness $\gtrsim$~20~mm has not been explored or reported thus far. As it is shown below, the good performance found here for the analytical approach becomes worse with increasing crystal thickness. This effect is particularly severe for the 30~mm thick crystal. In this respect, we have developed a methodology based on a $\chi^2$-discrimination approach, which allows one to recover a satisfactory gamma-ray hit localization at the cost of reconstruction efficiency. This method is described in more detail in the following sections.

\subsection*{\lacls 50$\times$50$\times$20~mm$^3$}
\begin{figure}[htbp!]
\flushleft
\centering
\includegraphics[width=\myfigsizelin\columnwidth]{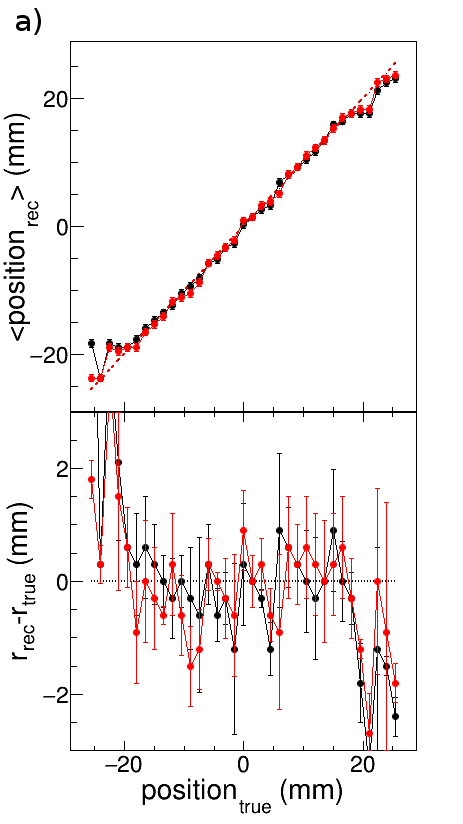}
\includegraphics[width=\myfigsizelin\columnwidth]{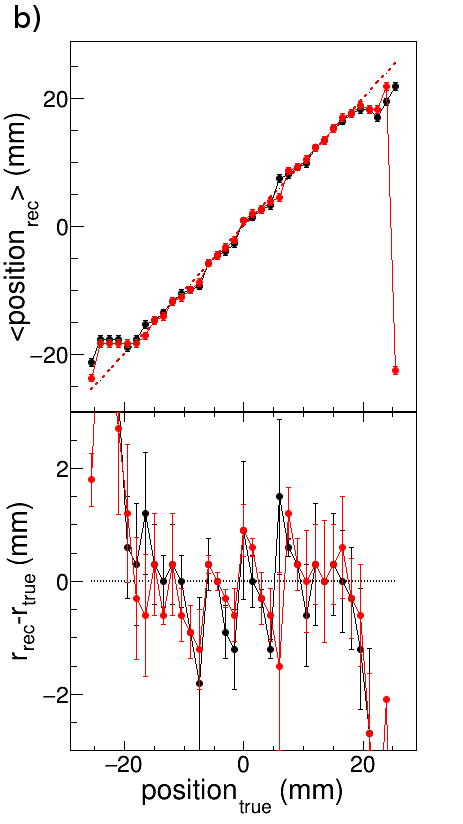}
\caption{\label{fig:linearity20mmAnalytical} Linearity along the diagonal scan positions in the 20~mm thick crystal obtained using the Lerche-fit (a) and Li-fit (b) methods.}
\end{figure}

\begin{figure}[htbp!]
\flushleft
\centering
\includegraphics[width=\myfigsize\columnwidth]{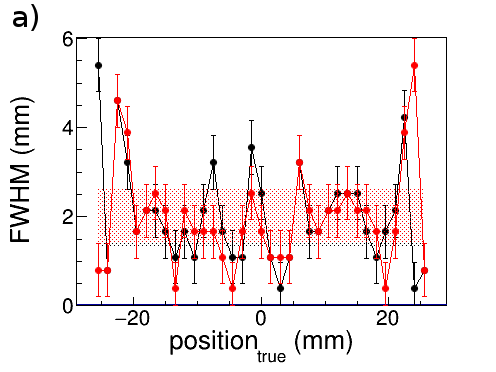}
\includegraphics[width=\myfigsize\columnwidth]{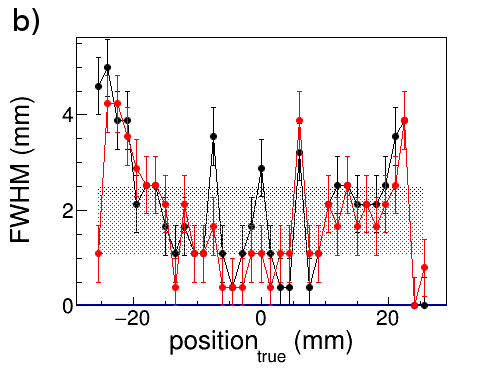}
\caption{\label{fig:resolution20mmAnalytical} Spatial resolution \textsc{fwhm} (mm) obtained for the 20~mm thick crystal using the Lerche- (a) and Li-model (b) fit methods.}
\end{figure}

Fig.~\ref{fig:linearity20mmAnalytical} shows the linearity curves obtained for the 20~mm thick crystal using both Lerche- and Li-model fit methods. The linearity performance found for both models is slightly worse than that found before for the 10~mm thick crystal. Fluctuations in linearity become now appreciably more pronounced. Border effects are also enhanced with respect to the 10~mm thick crystal, thus leading to FoVs of 14~cm$^2$ and 15.2~cm$^2$ for the Lerche- and Li-fit methods, respectively. The spatial resolution (Fig.~\ref{fig:resolution20mmAnalytical}) deteriorates slightly, with an average value of $\sim$2~mm \textsc{fwhm} for both models.

\begin{figure}[htbp!]
\flushleft
\centering
\includegraphics[width=\myfigsize\columnwidth]{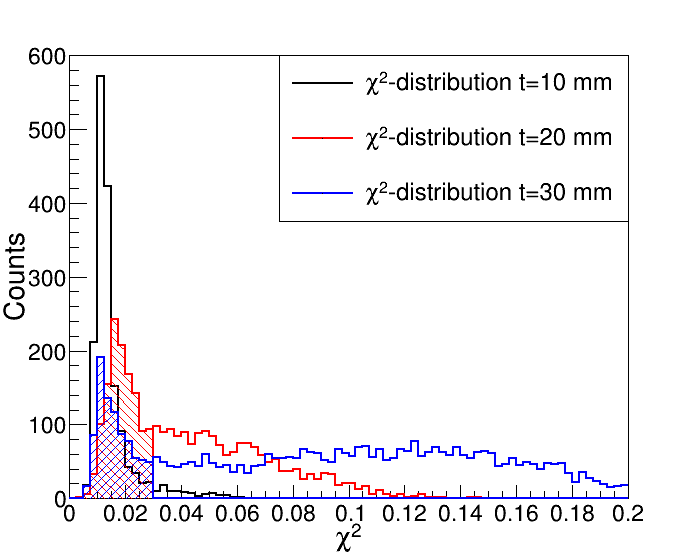}
\caption{\label{fig:chi2} $\chi^2$ distributions found for the Lerche-model applied to the three crystal thicknesses (see labels). The shaded region shows the selection made on the 20~mm and 30~mm thick crystals in order to preserve a position-reconstruction performance comparable to that of the 10~mm thick crystal.}
\end{figure}

The degradation of the analytical-model approach with increasing crystal thickness is demonstrated in Fig.~\ref{fig:chi2}, which shows the $\chi^2$-distribution from the Lerche-model fit for all three detectors investigated in this work. The decrease in the goodness of the fit with increasing crystal thickness may be related to the different aspect-ratio of the crystals, which has an impact on the characteristics of the scintillation-light propagation~\cite{Pauwels12}. Additionally, this degradataion may be also partially ascribed to the interplay between crystal thickness and the increasing contribution of multiple Compton hits eventually leading to full absorption. Both aspects seem to impact the light distribution in such a way, that it becomes more difficult for the analytical-model fit to properly identify the vertex of the main gamma-ray hit.

On the other hand, the $\chi^2$ value resulting from the fit can be used to circumvent precisely these problems by selecting events where the agreement between the model and the measured distribution is satisfactory (low $\chi^2$ values). This approach, however, implies a corresponding reduction on the overall reconstruction efficiency. In general, depending on the final detector application, a compromise can be chosen between reconstruction efficiency and position-localization accuracy.

\begin{figure}[htbp!] 
\flushleft
\centering
\includegraphics[width=\myfigsize\columnwidth]{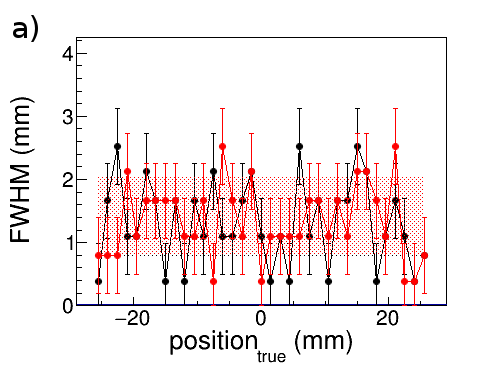}
\includegraphics[width=\myfigsize\columnwidth]{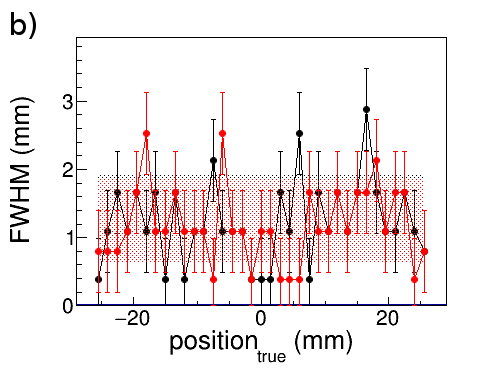}
\caption{\label{fig:resolution20mmAnalytical_Chi2cut} Same as Fig.~\ref{fig:resolution20mmAnalytical} using a selection of events on the corresponding $\chi^2$-distribution. See Fig.~\ref{fig:chi2} and text for details.}
\end{figure}

\begin{figure}[htbp!]
\flushleft
\centering
\includegraphics[width=\myfigsizelin\columnwidth]{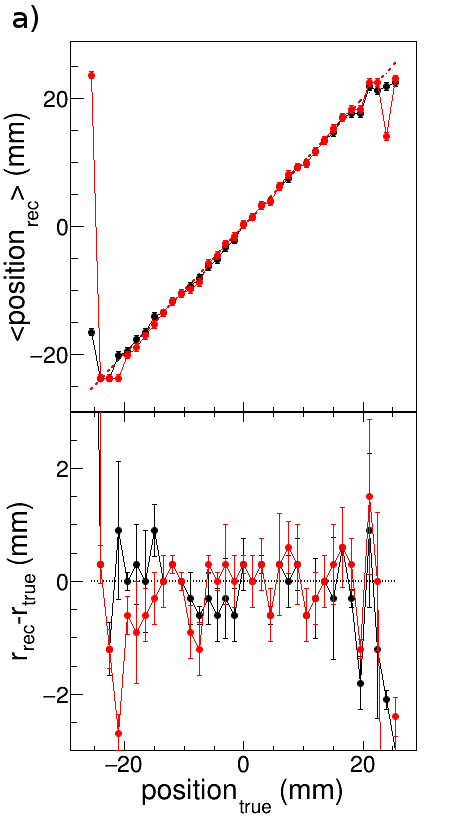}
\includegraphics[width=\myfigsizelin\columnwidth]{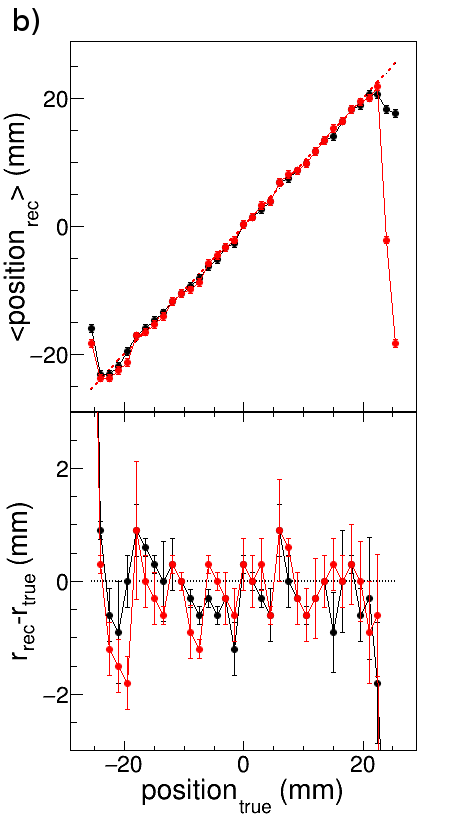}
\caption{\label{fig:linearity20mmAnalytical_Chi2cut} Same as Fig.~\ref{fig:linearity20mmAnalytical} using a selection of events on the corresponding $\chi^2$-distributions. See Fig.~\ref{fig:chi2} and text for details.}
\end{figure}

To illustrate this methodology for the 20~mm thick crystal we arbitrarily select events whose $\chi^2$-value is within the shaded region shown in Fig.~\ref{fig:chi2}. This $\chi^2$ range corresponds to nearly all events for the 10~mm thick crystal (where no $\chi^2$ selection was made), and represents about 40\% of the events in the 20~mm thick crystal. For this restricted data-set we obtain improved position resolutions (Fig.~\ref{fig:resolution20mmAnalytical_Chi2cut}), which are now comparable to those reported before for the 10~mm thick crystal. This selection on the $\chi^2$ distribution leads to an average position resolution of $\sim$1.3~mm (Fig.~\ref{fig:resolution20mmAnalytical_Chi2cut}). The linearity curves have less fluctuations and the FoV is also enhanced, when compared to the same data-set without $\chi^2$ selection (see Fig.~\ref{fig:linearity20mmAnalytical} and Fig.~\ref{fig:linearity20mmAnalytical_Chi2cut}). The FoV becomes 18.9~cm$^2$ and 21.6~cm$^2$ for the Lerche- and Li-models, respectively.


\subsection*{\lacls 50$\times$50$\times$30~mm$^3$}

\begin{figure}[htbp!]
\flushleft
\centering
\includegraphics[width=\myfigsizelin\columnwidth]{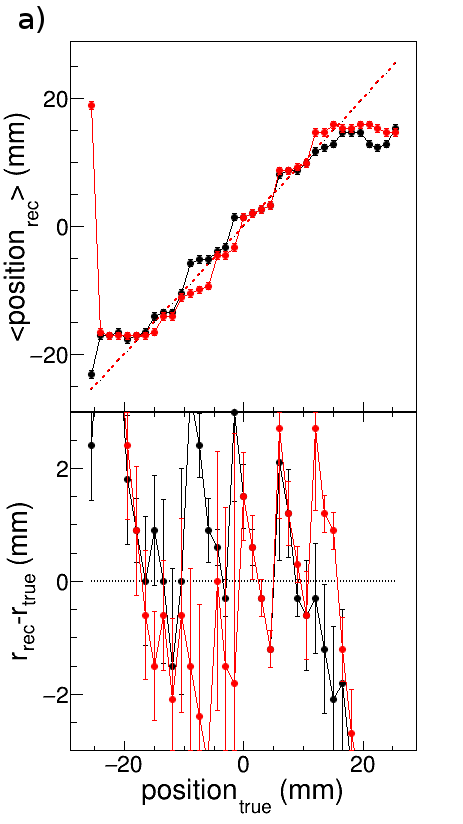} 
\includegraphics[width=\myfigsizelin\columnwidth]{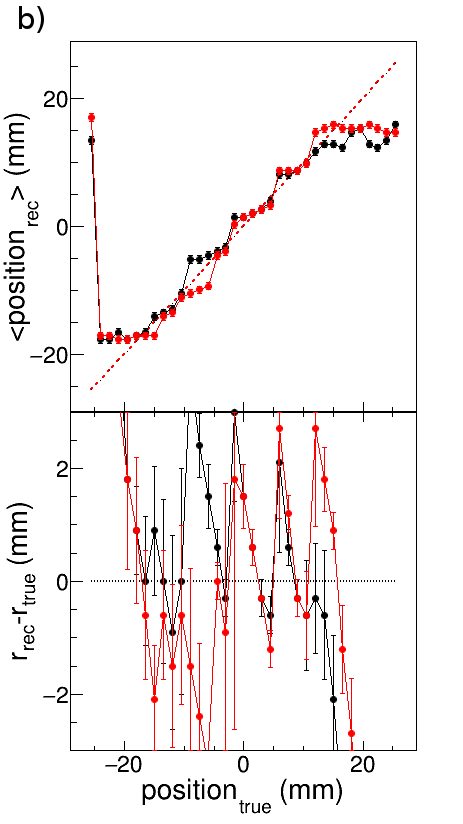}
\caption{\label{fig:linearity_fit_30mm} Linearity for the Lerche-fit (a) and Li-fit (b) methods applied to the 30~mm thick crystal without $\chi^2$-selection.}
\end{figure}

The shortcoming of the analytical-model approach becomes very apparent when applied directly to the 30~mm thick crystal. This statement is demonstrated in Fig.~\ref{fig:linearity_fit_30mm}, which shows the linearity obtained when an attempt is made to fit all registered events without any selection on the goodness of the fit. The average linearity fluctuations are of 1.6~mm \textsc{rms}. The position resolution becomes similar for both Lerche- and Li-model approaches, with an average value of $\sim$4~mm \textsc{fwhm}.

\begin{figure}[htbp!]
\flushleft
\centering
\includegraphics[width=\myfigsizelin\columnwidth]{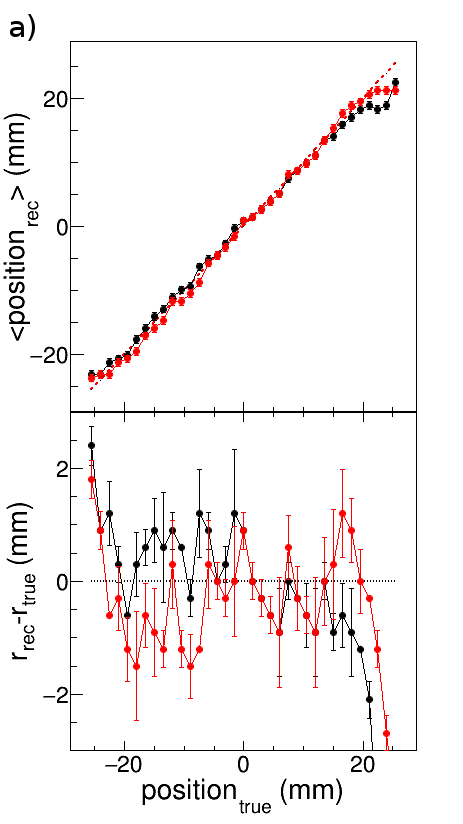}
\includegraphics[width=\myfigsizelin\columnwidth]{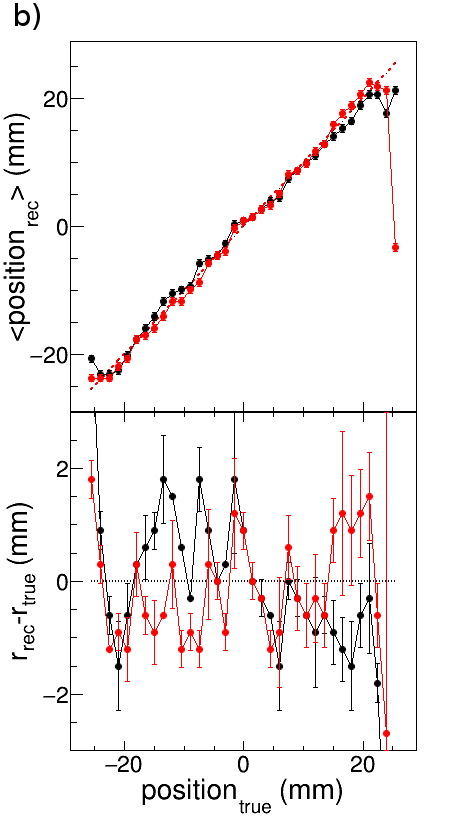}
\caption{\label{fig:linearity_fit_30mm_chi} Same as Fig.~\ref{fig:linearity_fit_30mm} with a $\chi^2$-based event selection. See also Fig.~\ref{fig:chi2} and text for details.}
\end{figure}

\begin{figure}[htbp!]
\flushleft
\centering
\includegraphics[width=\myfigsize\columnwidth]{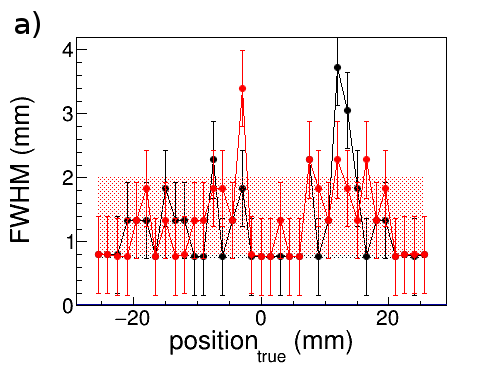}
\includegraphics[width=\myfigsize\columnwidth]{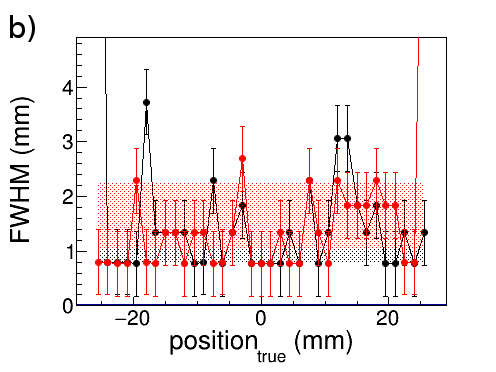}
\caption{\label{fig:resolution_fit_30mm} $\chi^2$-Gated spatial resolution obtained with the 30~mm thick crystal using the Lerche-model (a) and the Li-model (b).}
\end{figure}

Applying in the $\chi^2$-distribution of the 30~mm thick crystal the same selection that was used for the 20~mm thick crystal (see Fig.~\ref{fig:chi2}) one can recover, to some extent, a satisfactory position reconstruction. The $\chi^2$-gated linearity curves for the Lerche- and Li-model are displayed in Fig.~\ref{fig:linearity_fit_30mm_chi}. The linearity fluctuations are reduced now to an average value of $\sim$0.9~mm~\textsc{rms}. The improvement in performance can be appreciated by comparing Fig.~\ref{fig:linearity_fit_30mm} and Fig.~\ref{fig:linearity_fit_30mm_chi}. The new FoV becomes 21.6~cm$^2$ for both Lerche- and Li-fit methods. The $\chi^2$-gated position resolution (Fig.\ref{fig:resolution_fit_30mm}) is improved to an average level of $\sim$1.4~mm \textsc{fwhm} for both models. However, the $\chi^2$ selection becomes much more restrictive in terms of statistics owing to the large portion of events with relatively large $\chi^2$ values (see Fig.~\ref{fig:chi2}). Indeed, the $\chi^2$ selection for the 30~mm thick crystal represents only $\sim$15\% of the total statistics.

\subsection*{Summary of performances obtained with analytical-fit methods}

\begin{table*}[!htbp]
\caption{\label{tab:analytic}Summary of performances obtained with the two analytical models for the three different crystal thicknesses.}
\begin{center}
\begin{tabular}{cccccc}
  \hline
      &              & Resolution          & \textsc{rms}      &       & \\
Model &     Crystal size            &\textsc{<fwhm>}$_{x,y}$ & $r_{rec} - r_{true}$ & FoV &     S/N-Ratio      \\
      &     (mm$^3$)                    & (mm)                &     (mm)          & (cm$^2$)         &            \\
 \hline
       &50$\times$50$\times$10 & 1.20(15)  &     0.84(19)     &   15.2 &  14(3)\\
Lerche &50$\times$50$\times$20$^{(*)}$ & 1.24(10) &      0.69(8)      & 18.9 &   9(2)\\
       &50$\times$50$\times$30$^{(*)}$ & 1.32(20) & 0.86(13)     & 21.6 & 6(2)\\
\hline
   &50$\times$50$\times$10 & 1.24(10) & 0.86(23)& 18.9 &  12(5)\\
Li &50$\times$50$\times$20$^{(*)}$ & 1.46(12) & 0.67(4)  & 21.6 &  7(3)\\
   &50$\times$50$\times$30$^{(*)}$ & 1.43(12) & 0.88(16) & 21.6 & 4(2)\\
\hline
\end{tabular}
\newline\footnotesize{(*) With $\chi^2$-based event selection. See text for details.}
\end{center}
\end{table*}

The most remarkable feature of these methods is the attainable spatial resolution of 1.2-1.4~mm~\textsc{fwhm}. Interestingly, this position resolution can be achieved for all crystal thicknesses implementing only a minor previous characterization. Nevertheless, for thick crystals ($\gtrsim$20~mm) such an accuracy seems feasible only at a rather high cost in reconstruction efficiency. This aspect needs to be considered and evaluated for each particular detector application. In general, resolution, linearity and FoV show a remarkable improvement with respect to the Anger- and squared-charge centroiding approaches. In particular, the FoV remains practically constant using the Li-fit method, regardless of the crystal thickness, which may indicate the proper treatment of reflection effects within the model. For both models the S/N-ratio shows a systematic deterioration as a function of the crystal thickness. This feature also indicates that, generally speaking, the analytical model approach seems to be better suited for scintillation-crystals with thickness $\lesssim$10~mm.

It is worth noting that the rate-processing speed is more than a factor of two faster for the Lerche-model approach ($r_{Lerche-fit}=2433$~Events/s) than for the Li-method ($r_{Li-fit}=967$~Events/s). This can be ascribed to the simpler mathematical expression and lower number of variables. Such processing rates represent only 35\% and 14\% of the benchmark value $r_{Anger}$ (see Sec.~\ref{sec:centroid}), respectively.

\subsection{Artificial neural network algorithm}\label{sec:nn}
To implement a NN-algorithm for the position reconstruction we make use of the Multi-Layer-Perceptron class library (\verb TMultiLayerPerceptron ) of the CERN ROOT package~\cite{root}. From the different learning-methods available in this class, we find the quasi-Newton approach by Broyden-Fletcher-Goldfarb-Shanno (BFGS) to be the one providing best results. Similarly as reported in Ref.~\cite{Ulyanov17}, we also find noticeably better results when two decoupled and independently trained NNs are used, one for the $x$-coordinate and another one for the $y$-coordinate, rather than a single network with two outputs (x,y).

\begin{figure}[htbp!]
\flushleft
\centering
\includegraphics[width=\columnwidth]{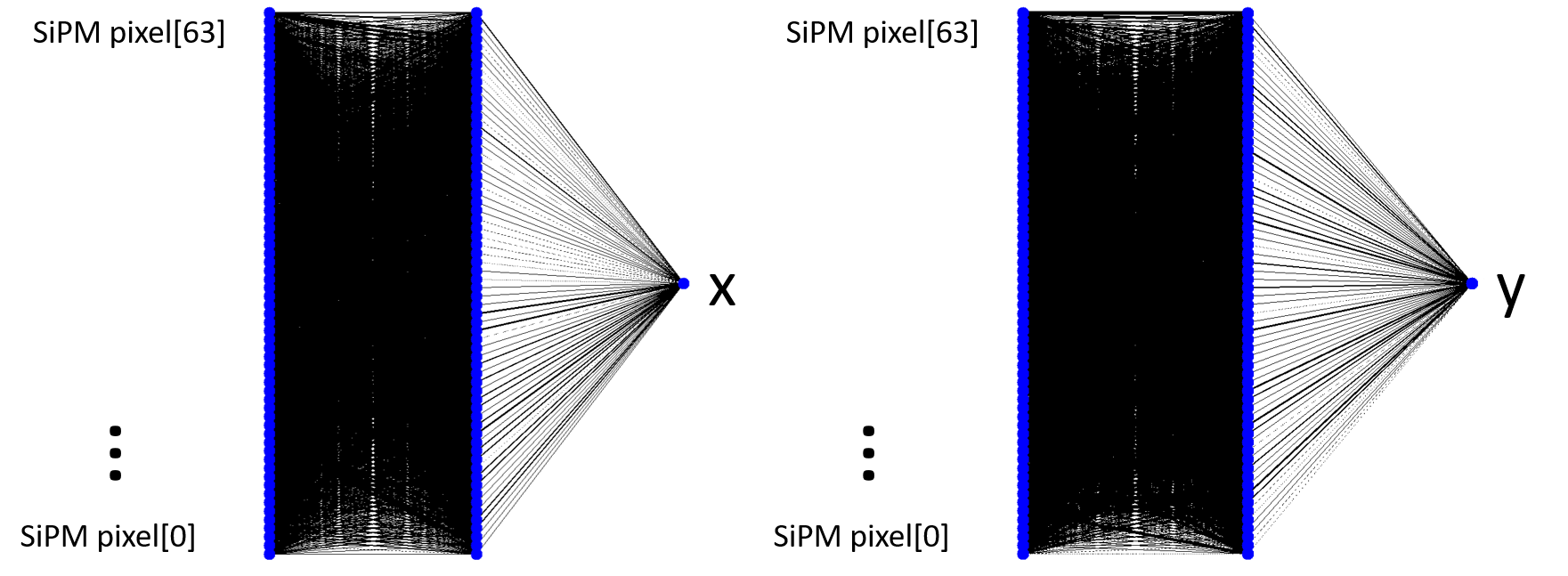}
\caption{\label{fig:nn} Diagram of two neural networks used in this work for the reconstruction of the $x$- and $y$-coordinates. Neurons are represented by bold-blue circles. In each NN the first (left-side) layer represents the 64 passive input neurons related to the 8$\times$8 pixels in the SiPM. The middle hidden-layer consists of 64 active neurons. The last layer is one single passive neuron for the output, which is either the $x$- or the $y$-coordinate for each NN. The strength of the lines represents the weight of the connections between neurons.}
\end{figure}
Fig.~\ref{fig:nn} shows an schematic diagram of two NNs used for the reconstruction of the $x$- and $y$-coordinates. With our 64-channel SiPMs, the NN input is always a 64 neurons passive-layer, which represents the charge-content of the 8$\times$8 matrix of SiPM pixels at each event. We have investigated different options for the nodes-structure of the intermediate NN layer and, in general, a single active layer of 64~neurons (as shown in Fig.~\ref{fig:nn}) seems to be the best approach in terms of linearity performance and accuracy. The last layer of the NN consists of just one output neuron ($x$ or $y$), which at the NN-training stage reperesents the corresponding coordinate for the scan position of the 35$\times$35 independent measurements used to train the network (see Sec.\ref{sec:setup}).
Depending on the crystal thickness, typically between 1.5$\times$10$^6$ and 3$\times$10$^6$ events are collected at each position of the 35$\times$35 characterization grid (see Sec.~\ref{sec:setup}). Half of the measured events are used for training the network and the other half are used for the iterative convergence test. With such statistics, and the aforementioned NN structure, approximately 6 minutes are required for each training cycle using a core-i7 processor from Intel.   Before performing the training of the NN the input data-base is pre-filtered by removing events whose charge distribution has a maximum, which is located at least at a distance beyond $\delta_{df} = 30$~mm from the maximum of the accumulated charge distribution. This allows one to make a more efficient use of the training resources and CPU time by excluding, a priori, a small fraction of stray events or random coincidences (typically 5-10\%). We use a total number of 150 epochs to train the NN in order to keep the overall calculation time within reasonable limits (about 15~h for each NN-training). Although generally the network has not fully converged after such a small number of cycles, the margin for further improvement with additional cycles seems rather negligible.
Corrections to account for pixel-gain fluctuations seem to play a minor role in NN-based methods, thus we found no difference between implementing or neglecting such experimental effects. This indicates that the NN seems capable to account itself quite reliably for gain-inhomogeneities along the learning process.
  
\subsection*{\lacls 50$\times$50$\times$10~mm$^3$}

\begin{figure}[htbp!]
\flushleft
\centering
\includegraphics[width=0.68\columnwidth]{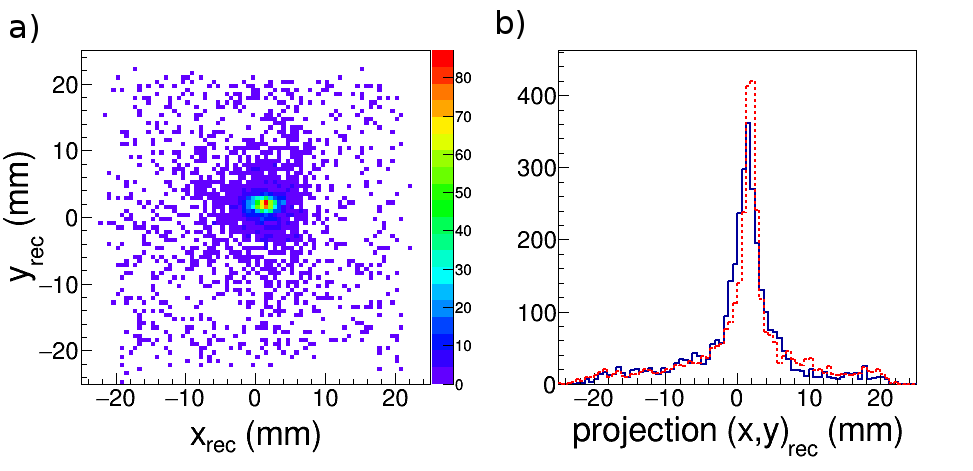}
\includegraphics[width=0.68\columnwidth]{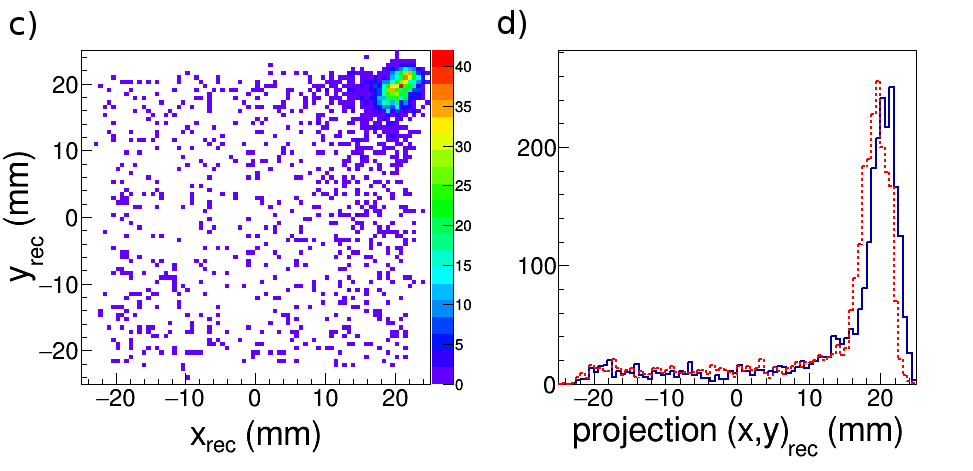}
\caption{\label{fig:nn10mm_examples} Examples of reconstructed positions with the NN-algorithm for the center of the crystal (a-b) and for a diagonal position shifted 21~mm in $x$ and $y$ (c-d) for the 10~mm thick crystal.}
\end{figure}

\begin{figure}[htbp!]
\flushleft
\centering
\includegraphics[width=\myfigsizelin\columnwidth]{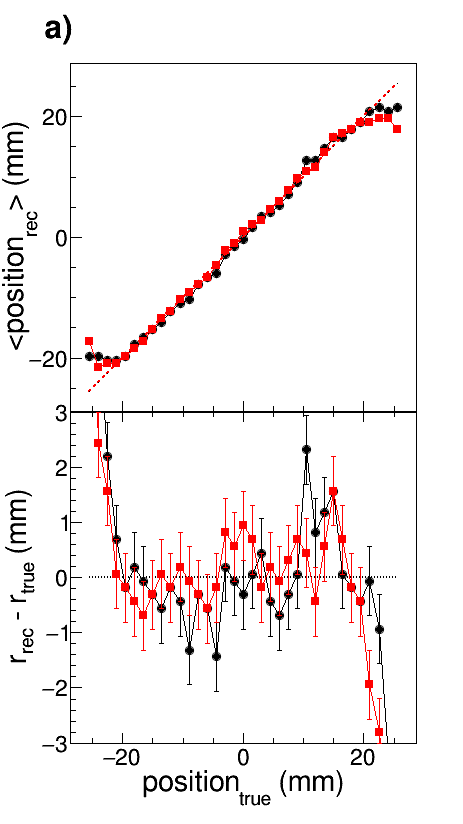}
\includegraphics[width=\myfigsizelin\columnwidth]{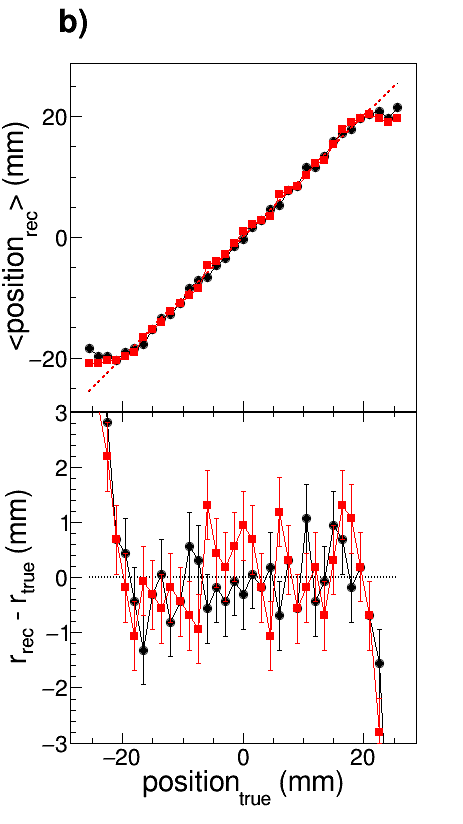}
\caption{\label{fig:nn10mm linearity} Linearity along the crystal diagonal (a) and horizontal-vertical cross (b) for the 10~mm thick crystal.}
\end{figure}

\begin{figure}[htbp!]
\flushleft
\centering
\includegraphics[width=0.68\columnwidth]{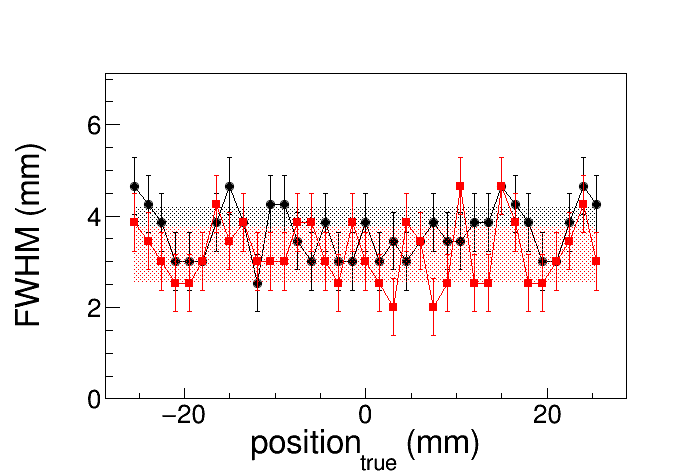}
\caption{\label{fig:nn10mm_fwhm} Spatial resolution (FWHM) obtained for the 10~mm crystal along the central $x$ (black) and $y$ axis (red). Shadow-bands indicate average values. }
\end{figure}

Fig.~\ref{fig:nn10mm_examples} shows the two reference illustrative examples for the 2D-position reconstruction when the NN-approach is applied to the 10~mm thick \lacls crystal. The first noticeable difference with respect to analytical methods is due to the broader spatial distributions obtained with the NN. In order to reliably assess the validity of the NN algorithm it becomes convenient to explore its performance along the full crystal surface. Indeed, for other NN-structures with lower number of active neurons than the one shown in Fig.\ref{fig:nn}, the results were satisfactory along e.g. the central crystal axis, but failed dramatically in other regions (diagonals) of the crystal. Thus, the overall validity of the chosen NN structure (Fig.~\ref{fig:nn}) is demonstrated by the linerarity curves of both the central-cross and diagonal-set of scanned positions along the crystal surface, which are shown in Fig.~\ref{fig:nn10mm linearity} for the 10~mm thick crystal.  
Similarly as it was found with analytical methods in the previous section, for NNs the improvement in FoV becomes very apparent when compared to the Anger-logic and squared-charge approaches shown in Sec.~\ref{sec:centroid}. The NN algorithm indeed yields a linear performance in a range of 46.5~mm both along the central $x$- and $y$-axis, as well as along the $x$- and $y$-projections of the crystal diagonal. This leads to an squared FoV of $\sim$21.6~cm$^2$. Within such FoV, the linearity slope is practically 100\% along any axis or diagonal. Sudden deviations of the linearity, such as the one occuring at $x=10$~mm for the diagonal data-set, lead to local maximum discrepancies of $\sim$2~mm. Most of the remaining discrepancies, both for the diagonal and for the central $x-$ and $y-$axis (Fig.~\ref{fig:nn10mm linearity}), are within about $\pm$1~mm, being the average value 0.85(5)~mm \textsc{rms}.
The spatial resolution is displayed in Fig.~\ref{fig:nn10mm_fwhm} for the central-cross of scanned positions along the $x-$ and $y-$axis. On average we find a resolution of 3.35(11)~mm~\textsc{fwhm} for the 10~mm thick crystal. The average S/N ratio becomes 12.0(2).

\subsection*{\lacls 50$\times$50$\times$20~mm$^3$}
\begin{figure}[htbp!]
\flushleft
\centering
\includegraphics[width=0.68\columnwidth]{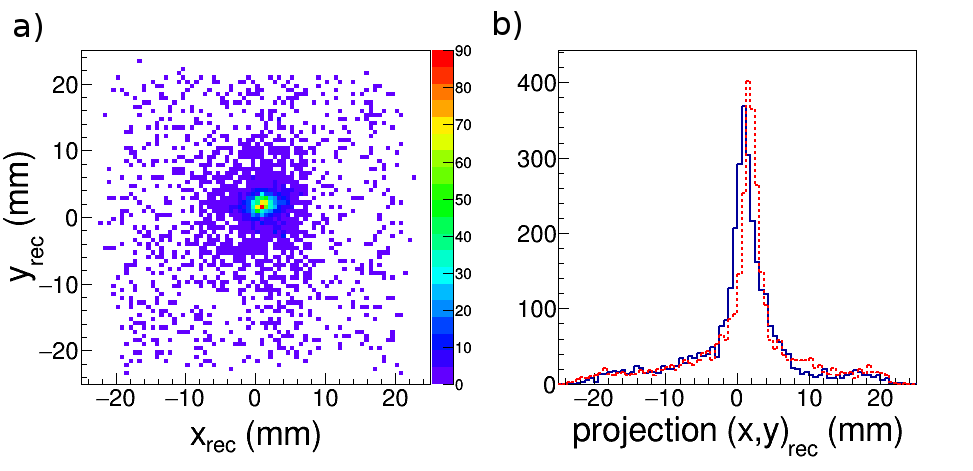}
\includegraphics[width=0.68\columnwidth]{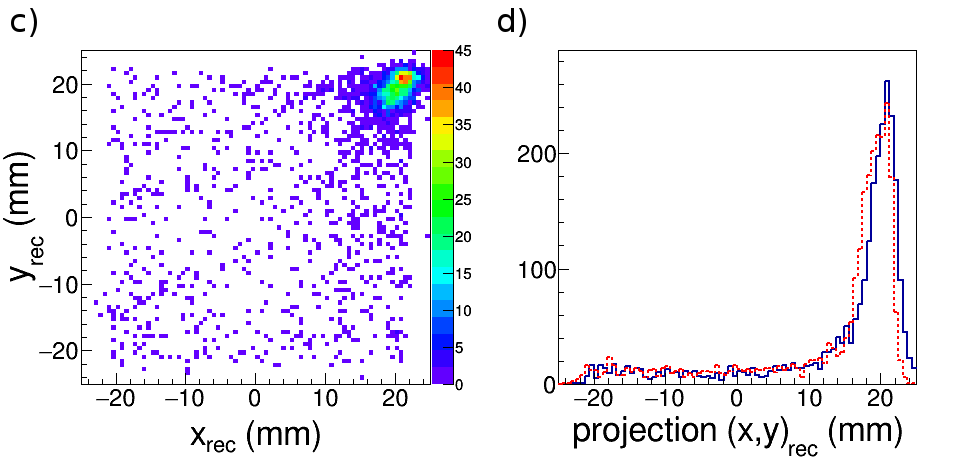}
\caption{\label{fig:nn20mm_examples} Examples of two position-distributions for the center of the crystal (a-b) and for a diagonal position shifted 21~mm in $x$ and $y$ (c-d) for the 20~mm thick crystal.}
\end{figure}

Using NNs the quality of the reconstructed 2D-distributions is quite similar to that obtained for the 10~mm thick crystal. This is demonstrated in Fig.~\ref{fig:nn20mm_examples}, which shows the same two scan positions of Fig.~\ref{fig:nn10mm_examples}, as measured and reconstructed now using a NN for the 20~mm thick crystal.

\begin{figure}[htbp!]
\flushleft
\centering
\includegraphics[width=\myfigsizelin\columnwidth]{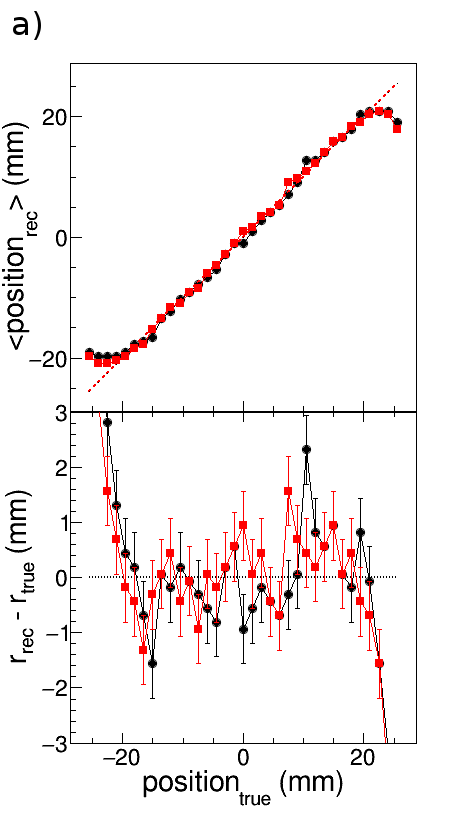}
\includegraphics[width=\myfigsizelin\columnwidth]{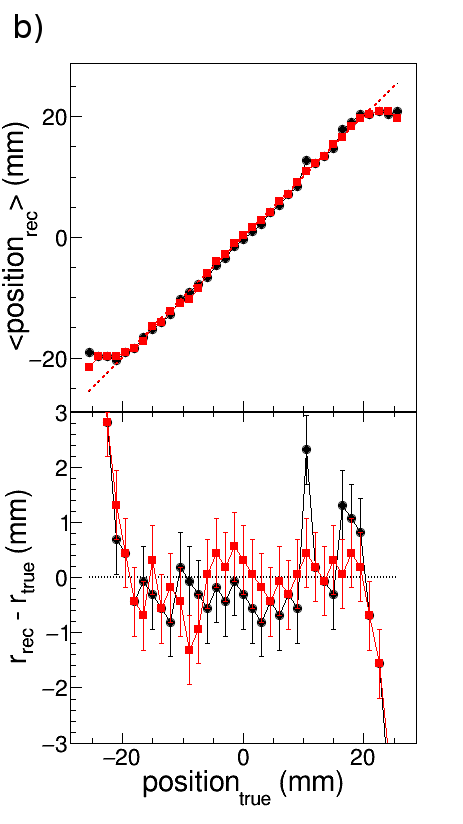}
\caption{\label{fig:nn20mm} Linearity for the NN-algorithm along the crystal diagonal (a) and the horizontal-vertical cross (b) in the 20~mm thick crystal.}
\end{figure}

The linearity curves for the 20~mm thick crystal are displayed in Fig.~\ref{fig:nn20mm} together with the discrepancies related to the ideal-detector performance. The linear range is ascribed to the central 31$\times$31 scanned positions, thus yielding also a FoV of 21.6~cm$^2$. The average deviation found is 0.83(6)~mm \textsc{rms}. 

\begin{figure}[htbp!]
\flushleft
\centering
\includegraphics[width=0.68\columnwidth]{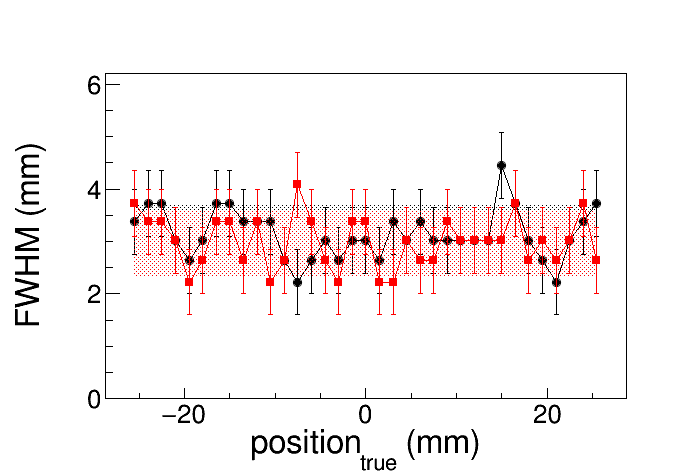}
\caption{\label{fig:nn20mm_fwhm} Spatial resolution (FWHM) obtained for the 20~mm crystal along the central $x$ (black) and $y$ axis (red) for the 20~mm thick crystal. Shadow-bands indicate average values. }
\end{figure}

The spatial resolution (Fig.~\ref{fig:nn20mm_fwhm}) and the S/N ratio become comparable or slightly better than for the 10~mm thick crystal, with average values of 3.01(11)~mm~\textsc{fwhm} and 12.3(4), respectively.

\subsection*{\lacls 50$\times$50$\times$30~mm$^3$}
\begin{figure}[htbp!]
\flushleft
\centering
\includegraphics[width=0.68\columnwidth]{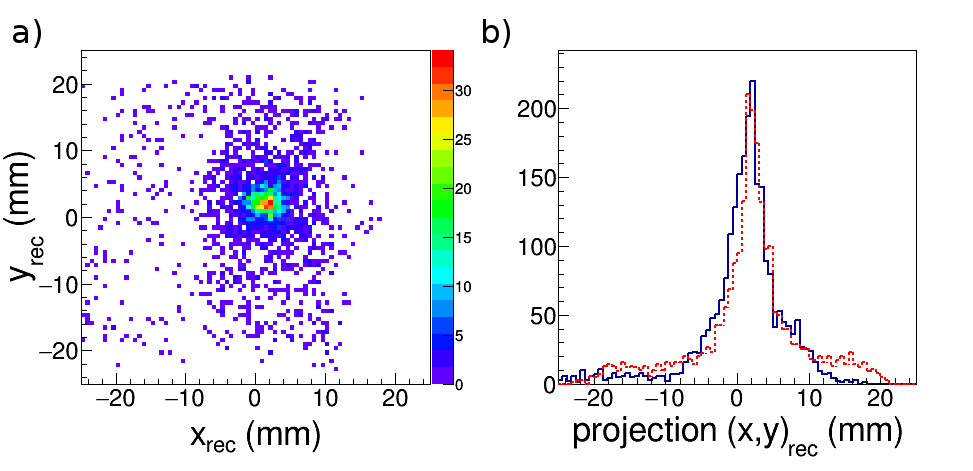}
\includegraphics[width=0.68\columnwidth]{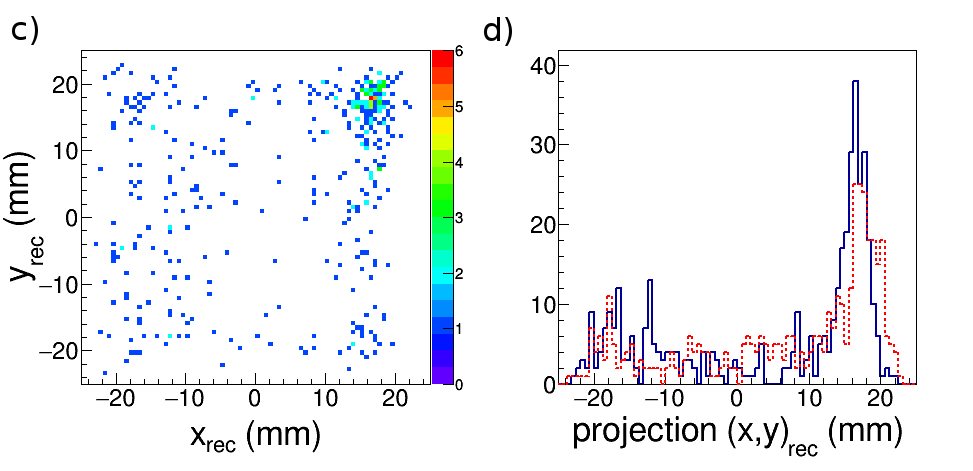}
\caption{\label{fig:nn30mm_examples} Examples of two position-distributions obtained using the NN-algorithm with the 30~mm thick crystal for the center of the crystal (a) and for the peripheral scan position (b).}
\end{figure}

\begin{figure}[htbp!]
\flushleft
\centering
\includegraphics[width=\myfigsizelin\columnwidth]{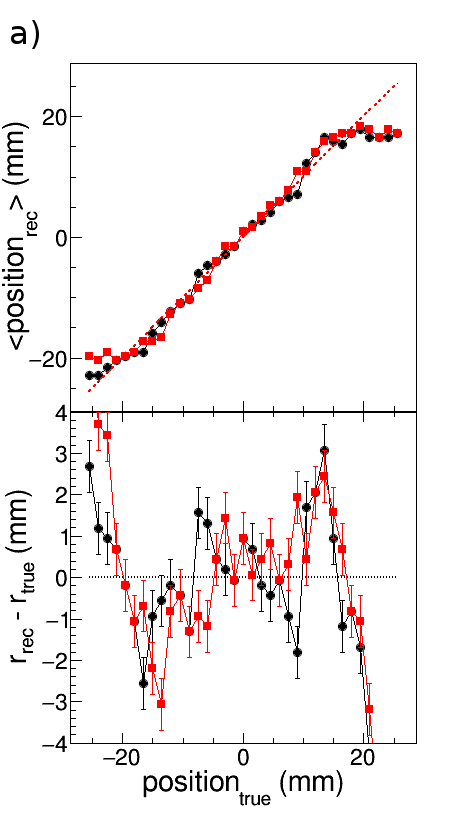}
\includegraphics[width=\myfigsizelin\columnwidth]{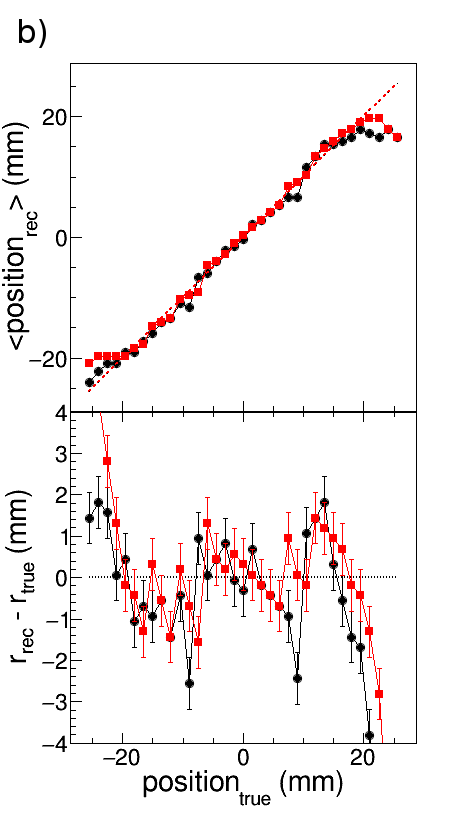}
\caption{\label{fig:nn30mm} Linearity for the NN-algorithm along the crystal diagonal (a) and central-axes (b) for the 30~mm thick crystal.}
\end{figure}

The mild differences between the 10~mm and 20~mm thick crystals become now more apparent for the 30~mm thick crystal. The two reference (central and peripheral) 2D-distributions are shown in Fig.~\ref{fig:nn30mm_examples}.
The most remarkable impact of the large crystal thickness is the enhanced border distortion, which can be observed in the linearity curves displayed in Fig.~\ref{fig:nn30mm}. In this case the FoV has to be limited to the central 28$\times$28 scanned positions (17.6~cm$^2$). Average deviations with respect to the true positions show an average \textsc{rms}-value of 1.1~mm.

\begin{figure}[htbp!]
\flushleft
\centering
\includegraphics[width=0.8\columnwidth]{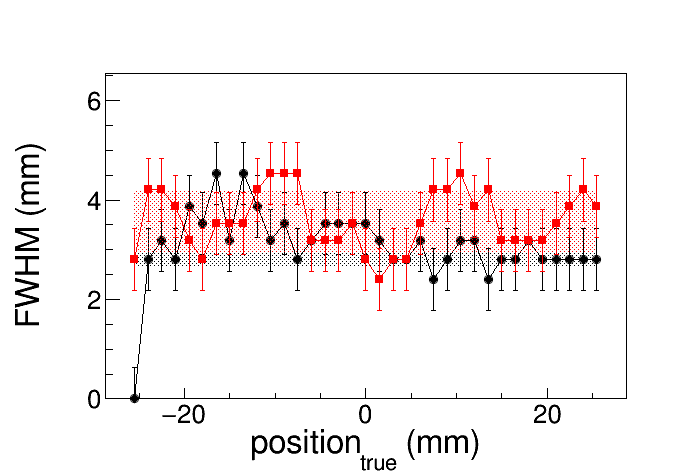}
\caption{\label{fig:nn30mm_fwhm} Spatial resolution (FWHM) obtained for the 30~mm crystal along the central $x$ (black) and $y$ axis (red). Shadow-bands indicate average values. }
\end{figure}

In the range of the FoV, the average spatial resolution becomes 3.36(15)~mm \textsc{fwhm} (see Fig.~\ref{fig:nn30mm_fwhm}) and the S/N ratio worsens to an average value of 7.0(4).

\subsection*{Summary of performances obtained with NN-algorithms}
\begin{table*}
\caption{\label{tab:nn}Summary of the performance results obtained with NN-based algorithms for the three different PSDs.}
\begin{center}
\begin{tabular}[width=\textwidth]{cccccc}
  \hline
  Crystal size          &  Resolution                   &   \textsc{rms}  &    FoV   & S/N-Ratio\\
  (mm$^3$)              & $< \textsc{fwhm}>_{(x,y)}$ (mm) &    $r_{rec} - r_{true}$     & (cm$^2$)  & \\
                        &                               &        (mm)          &           & \\
\hline
50$\times$50$\times$10 &  3.35(11)  &  0.86(7) & 21.6 & 12.0(2) \\
50$\times$50$\times$20 &  3.01(11)  &  0.83(6) & 21.6 & 12.3(4) \\
50$\times$50$\times$30 &  3.4(11)  &  0.94(16) & 17.6 & 7.0(4) \\
\hline
\end{tabular}
\end{center}
\end{table*}

The main performance parameters for the NN-based algorithms are summarized below in Table~\ref{tab:nn}. Using NNs the quality of the spatial reconstruction is rather similar for crystal thicknesses of 10~mm and 20~mm, with avereage position resolutions of $\sim$3~mm \textsc{fwhm} over the 21~cm$^2$ FoV. On the other hand, the 30~mm thick crystal shows a comparable performance in terms of resolution ($\sim$3.4~mm \textsc{fwhm}) but the FoV is reduced to 81\% of the field attainable with the thinner crystals. In terms of processing speed the NN-algorithm is rather fast (5200 Events/s). This represents $\sim$76\% of the Anger-algorithm processing rate $r_{Anger}$ (see Sec.~\ref{sec:centroid}). Other aspects about the performance found for the NN-algorithm will be discussed below in Sec.\ref{sec:summary} in the context of a comparison with the analytical methods.

\section{Depth of Interaction (DoI)}\label{sec:doi}
Initially we started to research a self-consistent approach, where all three space coordinates $x$, $y$ and $z$ were included in the position-reconstruction analysis, both for the analytical-fit and for the NN-based algorithms of Sec.~\ref{sec:fit} and Sec.~\ref{sec:nn}, respectively. However, we have found similar or better performances in both cases, when decoupling the position reconstruction in the transversal ($x,y$)-plane from the DoI analysis. To some extent, this may be related to the nature of the problem because a much higher-sensitivity and precision is expected for the $x,y$-coordinates than for $z$. In the former case variations on the first momentum of the distributions are very well estimated by the position of its maximum over the full detector surface. In the second case, however, only small changes in the second momentum are perceivable in the measured distribution, at least with the used instrumentation and SiPM pixelation.

\begin{figure}[htbp!]
\flushleft
\centering
\includegraphics[width=0.75\columnwidth]{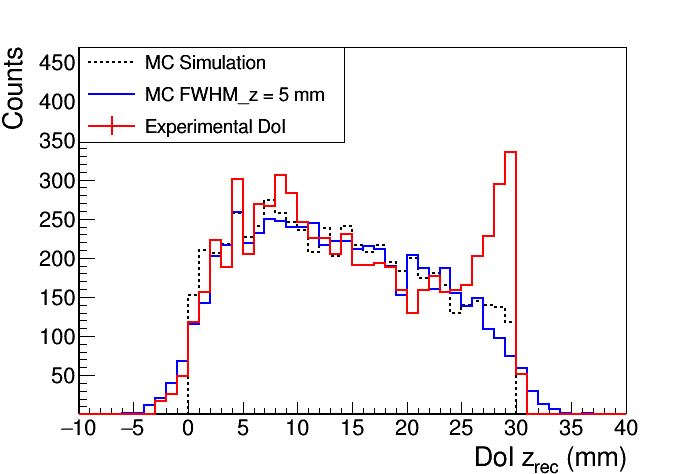}
\caption{\label{fig:mcdoi} DoI-calibrated from the measured area ($A_w$) of the scintillation-light distribubion at half maximum (red-line histogram). MC simulation of DoI values without broadening (dashed-line histogram) and with 5~mm \textsc{fwhm} broadening (blue-line histogram).}
\end{figure}

Thus, in order to determine the DoI for the \g-ray hit we use a rather phenomenological approach based on the inverse dependency of the $z$-coordinate (measured from the entrance surface) with the second momentum of the scintillation distribution. Neglecting perturbations induced by reflection effects we assume an inverse-linear relationship between DoI and the cross-section of the distribution $A_w$ at a given height $h_{w}$. This assumption is naturally expected from the parameterizations reported e.g. in Refs.\cite{Lerche05,Li10}. We choose $h_{w}$ the half-height value for each particular event. Smaller $h_w$ values are significantly affected by background light and spurius fluctuations. Values of $h_{w}$ closer to the maximum of the distribution make $A_w$ less senstive to the DoI because then $A_w$ starts to be dominated by the size of the pixel (6~mm).  In order to avoid artifacts in the reconstructed DoI arising from the 6~mm wide sampling resolution before computing $A_w$ for each measured event we perform a linear interpolation onto a 1~mm grid. Because DoI is more relevant for thick crystals, we focus here on the \lacls with size of $50\times50\times30$~mm$^3$. Our linear assumption for the DoI calibration is well justified, as demonstrated in Fig.~\ref{fig:mcdoi}, where the measured values for $A_w$ at half maximum (already calibrated) are compared against MC calculated DoIs. True or ideal simulated DoI values are shown by the dashed-line distribution in Fig.~\ref{fig:mcdoi}. The data used in this comparison corresponds to the central scan position $x=0$~mm and $y=0$~mm, where border effects can be safely neglected. The prominent peak in the experimental distribution arises from \g-ray hits near the optical window (large DoIs), where most of the charge above $h_w$ is concentrated in just one single pixel of the Si-PM. This leads to a slight overestimation of events with large DoI-values. The latter represent less than 10\% of the total measured events. This artifact could be reduced by lowering the value of $h_w$, nevertheless at the cost of higher incertitude on the estimated DoI.

In order to estimate an uncertainty for the experimentally determined DoI we compare its distribution with respect to the values obtained from a broadened MC-simulation. We find acceptable agreement between the measured $A_w$ distribution and the MC-simulated distribution for a Gaussian broadening of 5~mm~\textsc{fwhm}, which represents then a reasonable estimate for the uncertainty on the DoI.

For peripheral $\gamma$-ray hits in the crystal the width and the shape of the scintillation-light distribution depends not only on the DoI, but also on reflection effects. Thus, we use the 35$\times$35 scanned positions (see Sec.~\ref{sec:setup}) in order to determine, at each of them, i.e. on a grid of 1.5~mm$\times$1.5~mm the $A_w$ values for the broadest and narrowest light distributions at half maximum, which are assigned to DoI~=~0~mm and DoI~=30~mm, respectively. A linear regression is calculated for each scan position in order to interpolate any intermediate value for the DoI. DoI reconstruction examples for scan positions at the center and at the corner of the 30~mm thick crystal are shown below in Fig.~\ref{fig:doi}. 

\begin{figure}[htbp!]
\flushleft
\centering
\includegraphics[width=0.9\columnwidth]{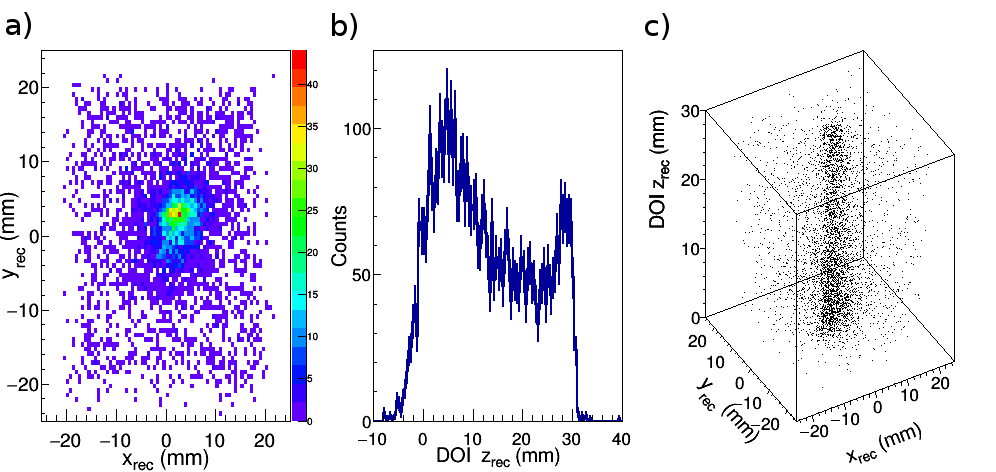}
\includegraphics[width=0.9\columnwidth]{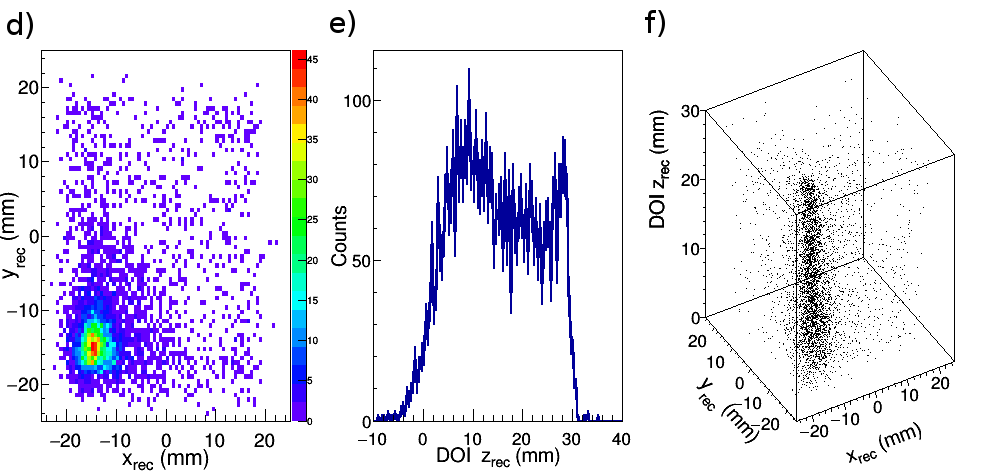}
\caption{\label{fig:doi} Examples of reconstructed DoI coordinates for the central scan position (a-c) and for a peripheral scan position (d-f). In each case the 2D ($x,y$) positions as reconstructed using a NN-algorithm are shown (a,d) and the corresponding calibrated DoI distributions are shown in panels (b,e). The distribution of 3D-coordinates is shown in (c,f).}
\end{figure}

The list of 35$\times$35 linear-regression coefficients are then stored in a single file for each crystal thickness. For an arbitrary measurement, the corresponding DoI-calibration coefficients are invoked after the ($x,y$) coordinates have been determined either using the analytical or the NN-approach.
In summary, one can conclude that this is a rather simple, yet reliable approach for determining the DoI at each ($x,y$)-coordinate within the FoV of the crystal, with an uncertainty of $\sim$5~mm~\textsc{fwhm}. Similar results are obtained for the crystals with thicknesses of 10~mm and 20~mm.

\section{Summary and outlook}\label{sec:summary}

\begin{figure}[htbp!]
\flushleft
\centering
\includegraphics[width=0.9\columnwidth]{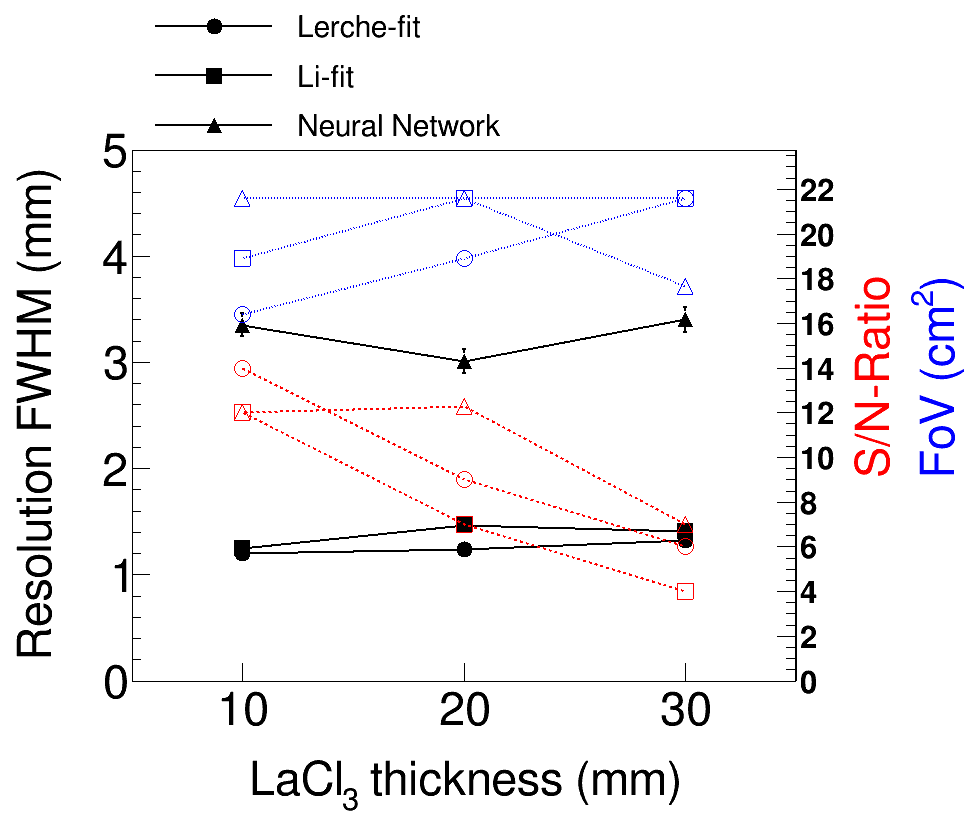}
\caption{\label{fig:summary} Solid symbols represent the spatial resolution (left-vertical axis) achieved with the different position reconstruction techniques (see legend) as a function of the crystal thickness (horizontal axis). Empty symbols represent (right-vertical axis) average S/N-Ratio (red) and FoV (blue).}
\end{figure}

The main performance features found for the different position-reconstruction algorithms have been summarized in Table~\ref{tab:analytic} and Table~\ref{tab:nn}. These results are graphically displayed in Fig.~\ref{fig:summary} for comparison.

An interesting result which derives from this study is that, for both analytical-model fit (with $\chi^2$-selection) and NN-based approaches, the crystal thickness has a minor impact on the position resolution, whereas its effect is more sizeable in  terms of linearity, FoV and S/N-ratio. Regarding spatial resolution the anlytical-fit methods show a superior performance (on average $\sim$2 times better than NN-algorithms), however at the cost of reconstruction efficiency for the thick crystals ($\gtrsim$20~mm).
On the other hand, the NN-algorithm shows a robust performance in terms of linearity and FoV, becoming the attainable spatial resolution its main limitation ($\gtrsim$3~mm \textsc{fwhm}). In summary, at least for parallelepiped crystals with a base size of 50$\times$50~mm$^2$, one can conclude that analytical methods are well suited for thin crystals ($\lesssim$10~mm), whereas NN-based algorithms may be a better choice for thick crystals ($\gtrsim$20~mm).

The spatial resolution obtained here for the analytical-fit methods applied to the 10~mm thick crystal ($\sim$1.2~mm \textsc{fwhm}) is comparable to the results reported by Ling et al.~\cite{Ling08} and Li et al.~\cite{Li10} using similar crystal thicknesses of 8~mm and 10~mm, respectively. About the applicability and performance of analytical methods with thick scintillation crystals ($\gtrsim$20~mm) there was no information reported thus far in the literature. In this case we have shown that a similar performance in position reconstruction can still be achieved by implementing a discrimination of events based on the $\chi^2$-goodness of the fit.

Regarding NN-algorithms our linearity and resolution results are rather constant regardless of crystal thickness. Thus, the spatial resolution found here for the 10~mm thick crystal ($\sim$3.3~mm \textsc{fwhm}) is comparable to the 2.9~mm~\textsc{fwhm} value reported by Ulyanov et al.~\cite{Ulyanov17} using CeBr$_3$ crystals of smaller size (25$\times$25$\times$10~mm$^3$). This result is significantly better than the value of $\gtrsim$4.7~mm reported in Ref.~\cite{Gostojic16} for LaBr$_3$-crystals of 50$\times$50$\times$10~mm$^3$ volume. This may be due to the fact that in the latter work simulated detector responses were used to train the NN, at variance with the experimental approach followed here and in Ref.~\cite{Ulyanov17}.  For the 20~mm thick crystal, our result for the spatial resolution using NNs ($\sim$3~mm~\textsc{fwhm}) is significantly better than the 8~mm~\textsc{fwhm} value reported in Ref.~\cite{Ulyanov17} using LaBr$_3$(Ce) crystals of 28$\times$28$\times$20~mm$^3$ size. This difference may be due to the rather thick optical window (5~mm) used in the LaBr$_3$(Ce)-detector of the latter study, given that the NN-methodology implemented was rather similar in both studies. Finally, we have not been able to find any previous position-characterization study involving 30~mm thick scintillation crystals and NNs. In this respect our results confirm the applicability of NNs to monolithic crystals of this geometry without a remarkable degradation on performance.

With respect to the applicability of these results in the field of neutron capture measurements, and in particular in the framework of the HYMNS project, the loss of reconstruction efficiency by the analytical-fit methods seems to be a major drawback for their use in the second detection layer (absorber) of i-TED. This limitation can be fully circumvented by implementing a NN-algorithm for the position reconstruction in the thick scintillation crystals. As demonstrated here, NN algorithms show a similar FoV and, on average, better S/N ratios than analytical methods. Furthermore, the $\sim$3~mm \textsc{fwhm} spatial resolution attainable with NN-algorithms does not seem a limiting factor in terms of the proposed Compton-technique for background rejection, given that the related uncertainty on the Compton angle is still dominated by the energy resolution of \lacls crystals.


\section*{Acknowledgment}
This work has received funding from the European Research Council (ERC) under the European Union's Horizon 2020 research and innovation programme (ERC Consolidator Grant project HYMNS, with grant agreement nr. 681740). The authors acknowledge the Spanish Ministerio de Ciencia e Innovaci\'on under grants FPA2014-52823-C2-1-P, FPA2017-83946-C2-1-P and the program Severo Ochoa (SEV-2014-0398) for support. CDP acknowleges CSIC funding from PIE-201750I026. The are grateful to PETSys Electronics S.A. for technical support and I3M at UPV-Valencia for helpful collaboration. Finally, we would like to thank an anonymous referee, who helped us to improve some important aspects of this article.

\bibliography{bibliography}

\end{document}